\documentclass[twoside,12pt]{article}
\usepackage{epsfig}

\newcommand{\gsim}{\raisebox{-4pt}{$\,\stackrel{\textstyle
                                                         >}{\sim}\,$}}
\newcommand{\nn}{\nonumber}
\newcommand{\be}{\begin{equation}}
\newcommand{\ee}{\end{equation}}
\newcommand{\ba}{\begin{eqnarray}}
\newcommand{\ea}{\end{eqnarray}}
\newcommand{\req}[1]{(\ref{#1})}
\def\={\,=\,}
\newcommand{\ci}[1]{\cite{#1}}
\newcommand{\midtilde}{\raisebox{-0.25\baselineskip}{\textasciitilde}}

\def\gev{~{\rm GeV}}

\def\eps{\epsilon}

\newcommand{\tw}{\textwidth}
\def\vk{{\bf k}_{\perp}}

\def\vbs{{\bf b}}
\def\vb0{{\bf b}_0}

\def\={\,=\,}

\begin{document}

\title{Deeply Virtual Meson Production on the nucleon}

\author{L. Favart$^1$, M. Guidal$^2$, T. Horn$^{3,4}$, P. Kroll$^4$\\
\textit{$^1$IIHE - C.P. 230, Universit\'e Libre de Bruxelles}\\
\textit{1050 Brussels, Belgium}\\
\textit{$^2$Institut de Physique Nucl\'eaire d'Orsay}\\
\textit{CNRS-IN2P3, Universit\'e Paris-Saclay, 91406 Orsay, France.}\\
\textit{$^3$Thomas Jefferson National Accelerator Facility}\\ 
\textit{Newport News, Virginia 23606, USA}\\
\textit{$^4$The Catholic University of America}\\
\textit{Washington DC 20064, USA}\\
\textit{$^5$Fachbereich Physik, Universit\"at Wuppertal}\\
\textit{42097 Wuppertal, Germany.}\\}


\maketitle

\begin{abstract}
We review the status of the data and phenomenology in the Generalized Parton Distribution approach of Deep Virtual Meson Production.
\end{abstract}

\section{Introduction}
\label{sec:intro}

\begin{figure}[htb!]
\includegraphics[width=0.7\tw]{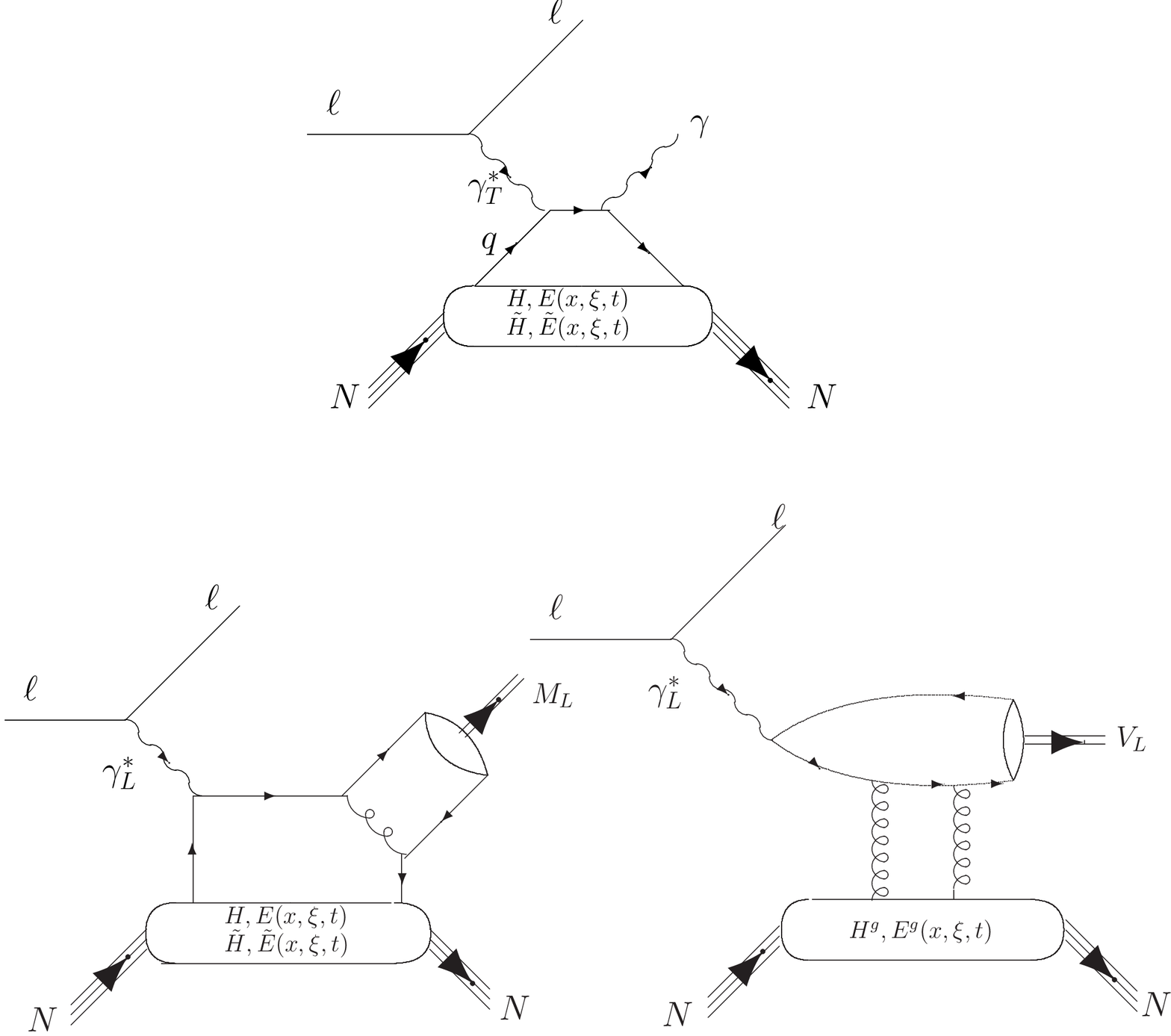}
\vskip -6.cm
\caption{Top: a typical leading-twist DVCS handbag diagram. Bottom: typical leading-twist
DVMP diagrams for the quark ($\gamma^*_Lq\to M_Lq$) and gluon ($\gamma^*_Lg\to V_Lg$)
subprocesses. The symbol $M$ stands for pseudo-scalar mesons as well as vector mesons
while the symbol $V$ stands only for vector mesons. The subscript $L$ ($T$) stands for
longitudinal (transverse) polarization.}
\label{fig:diags}
\end{figure}

In recent years hard exclusive processes like deeply virtual photon (DVCS) and 
meson (DVMP) leptoproduction have attracted much interest. It has been shown that in the 
generalized Bjorken regime of large photon virtuality, $Q^2$, and large energy, 
$W$, in the virtual photon - proton center of mass frame, but fixed Bjorken-$x$, 
$x_B$, these processes factorize in hard partonic subprocesses and soft hadronic 
matrix elements parametrized as the generalized parton distributions (GPDs) 
\ci{mueller,ji97,radyushkin96,radyushkin97}. Fig.~\ref{fig:diags} illustrates this factorization
for DVCS and DVMP. Rigorous proofs of factorization for DVMP are given 
in \ci{collins96,collins98}. In these references, it has also been shown that the asymptotically
dominant contribution comes from longitudinally polarized photons while those arising
from transversally
polarized photons are suppressed by $1/Q^2$. 
It is not clear that there is such factorization for transversally polarized photons.
For instance, in collinear approximation, the convolutions in the transverse 
amplitudes are infrared singular for light mesons.
However, in phenomenologically studies, factorization of the latter amplitudes has been
occasionally assumed to hold at least to leading-order of perturbation theory 
\ci{GK1,GK3,martin-ryskin-teubner}. Factorization in hard subprocesses and GPDs 
has also been shown to hold to next-to-leading order (NLO) of perturbative QCD 
for photo- (and low $Q^2$) leptoproduction of Quarkonia in the formal
limit of $m_Q\to\infty$ where $m_Q$ is the mass of the heavy quark~\cite{Ivanov:2004vd}.
In photoproduction, the photons are transversally polarized. In
low $Q^2$ leptoproduction of Quarkonia, longitudinal photons contribute,
but the photon polarization is still predominantly transverse.

DVMP is complementary to DVCS. The quantum numbers of the
produced meson allow one to probe different flavor combinations and allow
for disentangling the various flavor GPDs. DVMP also provides a way
to access transversity GPDs. However, the abundance of opportunities
provided by the produced meson in DVMP comes with additional experimental
and theoretical challenges. In contrast to DVCS, the leading-twist hard 
scattering subprocesses for DVMP
contain the exchange of hard quarks and gluons, which are accompanied by the appearance
of the strong coupling constant $\alpha_s$, and
a meson distribution amplitude, whose functional form is not completely
understood to date. As we will further discuss below, currently available
experimental DVMP data are not yet in the regime where the leading-twist
applies and strong power corrections are needed for their interpretation.


The present article focuses on DVMP and aims at reviewing the status 
of the data and phenomenology in the field. We start 
by recalling some particular aspects of the handbag approach and
the GPD phenomenology. For a detailed introduction to the theory of DVMP, 
we refer the reader to Refs.\ci{diehl04,rad-bel}
as well as to the introduction article devoted to GPDs and TMDs 
in this topical issue\ci{Diehl:2015uka}.
Then we will review the existing data for vector mesons and pseudoscalar mesons
and discuss their current interpretation in terms of GPDs.
We will conclude with a summary and the prospects for the field.



\section{DVMP and GPDs}

\subsection{DVMP handbag amplitudes}
\label{sec:amps}

The leading-twist DVMP amplitudes $\cal M$
are given by convolutions of the GPDs $K=K(x,\xi,t)$,
where $K$ is one of the leading-twist GPDs $H, E,\widetilde{H}, \widetilde{E}$ 
and the subprocess amplitudes, ${\cal H}$:
 
\ba
{\cal M}^{M(q)}_{0+,0+} &=& \frac{e_0}2 \sqrt{1-\xi^2}\sum_qe_q{\cal C}_M^q\int_{-1}^1 dx 
                     \sum_{\lambda} {\cal H}^{M(q)}_{0\lambda,0\lambda} \,
                       \Big[H^q_{\rm eff}+2\lambda \widetilde{H}^q_{\rm eff}\Big]\,,\nn\\
{\cal M}^{M(q)}_{0-,0+} &=& \frac{e_0}2 \mbox{sgn}(\Delta^1)
\frac{\sqrt{t_{\rm min}-t}}{2m}\sum_qe_q{\cal C}_M^q
  \int_{-1}^1 dx  \sum_{\lambda} {\cal H}^{M(q)}_{0\lambda,0\lambda} \,
                       \Big[E^q+2\lambda \xi\widetilde{E}^q\Big]\,, \nn\\
{\cal M}^{V(g)}_{0+,0+} &=& \frac{e_0}2 \sqrt{1-\xi^2}\sum_qe_q{\cal C}_V^q\int_{-1}^1 dx 
                     \sum_{\lambda} {\cal H}^{V(g)}_{0\lambda,0\lambda}\, 
                       H^g_{\rm eff}\,,\nn\\
{\cal M}^{V(g)}_{0-,0+} &=& \frac{e_0}2 \mbox{sgn}(\Delta^1)
\frac{\sqrt{t_{\rm min}-t}}{2m}\sum_qe_q{\cal C}_V^q
  \int_{-1}^1 dx  \sum_{\lambda} {\cal H}^{V(g)}_{0\lambda,0\lambda}\, E^g\,.
\label{eq:amplitudes}
\ea
Here, $m$ denotes the mass of the proton and the skewness, $\xi$, is related to 
Bjorken-$x$, $x_B$, by $\xi=x_B/(2-x_B)$ up to corrections of order $Q^2$.
$\Delta$ is the momentum transfer and $t=-\Delta^2$.
Terms of order $t/Q^2$ are usually neglected.
Forward scattering occurs at $t_{\rm min}=-4m^2\xi^2/(1-\xi^2)$.
In Eq.~\ref{eq:amplitudes}, the indices of $\cal M$
refer to the helicities of the initial and final protons ($\pm$) and 
to the helicity of the initial photon and final meson ($0$'s). The $\lambda$ indices
of $\cal H$ refer to the helicities of the partons participating
to the subprocess. While the four GPDs $H, E,\widetilde{H}, \widetilde{E}$ 
are all helicity-conserving at the parton level, one sees that $E$ and
$\widetilde{E}$ contribute to the nucleon helicity-flip amplitudes
and $H$ and $\widetilde{H}$ to the nucleon helicity-conserving amplitudes.
The symbols $q$ refer to the quark flavors, $g$ to gluons and ``sgn"
is the standard sign mathematical function. The non-zero flavor weight factors $\cal C$
are, for the electrically uncharged mesons like $\rho^0, \phi, \omega, \pi^0$:
\be
{\cal C}^u_{\rho^0, \pi^0}\=-{\cal C}^d_{\rho^0, \pi^0}\={\cal C}^u_\omega\={\cal C}^d_\omega\=1/\sqrt{2}\,,
\quad {\cal C}^s_\phi=1\,.
\ee
The generalization to other mesons is straightforward \ci{frankfurt99}.

Since the parton helicity, $\lambda$, is not observed it is to be summed over. The GPDs $H_{\rm eff}$ and $\widetilde{H}_{\rm eff}$
for quarks and gluons are the combinations
\be
H_{\rm eff}\=H - \frac{\xi^2}{1-\xi^2} E\,, \hspace*{0.1\tw}
\widetilde{H}_{\rm eff}\=\widetilde{H} - \frac{\xi^2}{1-\xi^2} \widetilde{E}\,.
\ee
The subprocess amplitudes, ${\cal H}$, for quarks and gluons, are to be calculated
perturbatively from an appropriate set of Feynman graphs. Typical leading-order
graphs for meson production are shown in the bottom two plots of Fig.\ \ref{fig:diags}.
For the quark subprocess amplitude they lead to
\be
\sum {\cal H}_{0\lambda,0\lambda}\=\frac{32}{9Q}\pi\alpha_s(\mu_R) f_M\langle 1/\tau \rangle_M\,
                              \Big[\frac1{x-\xi+i\eps}+\frac1{x+\xi-i\eps}\Big]\,,
\label{eq:leading-twist}
\ee
where $f_M$ and $\langle 1/\tau \rangle_M$ denotes the meson decay constant and the 
$1/\tau$-moment of its distribution amplitude, $\Phi_M(\tau)$; $\mu_R$ is an appropriate
renormalization scale of order $Q^2$. Analogous expressions hold for the gluonic subprocess.

We note the model-independent feature that the leading-twist handbag 
approach predicts a $1/Q^6$ scaling of the longitudinal cross section, 
$\sigma_L$, for DVMP at fixed $x_B$ which is modified by logarithms
from the evolution and the running of $\alpha_S$. The transverse 
cross section, $\sigma_T$, is expected to follow a $1/Q^8$ dependence
at fixed $Q^2$ and $x_B$.



\subsection{The parametrization of the GPDs}

Since the GPDs are controlled by non-perturbative QCD there is no analytic method known for 
their calculation. Lattice QCD only allows for the calculation of the lowest few moments of 
the GPDs at unphysical values of the quark masses \ci{haegler}. The extrapolation to the chiral 
limit is not fully understood \ci{bali,green}. Thus, in order to learn about 
them one tries to 
extract them from suitable experimental observables
like it is done for the parton density functions
(PDFs). Such extractions require parametrizations which are frequently constructed from the double distribution representation of the GPDs 
\ci{mueller,radyushkin99} assuming that the double distribution consists of a product of a
zero-skewness GPD and a weight function:
\ba
K(x,\xi,t)&=&\int_{-1}^1\,d\rho\,\int_{-1+|\rho|}^{1-|\rho|}\, 
                    d\eta\,\delta(\rho+\xi\eta-x)\,K(\rho,\xi=0,t)w(\rho,\eta)\nn\\
     && +D(x/\xi,t)\,\Theta(\xi^2-x^2)\,.
\label{eq:theory-DD}
\ea

We refer the reader to Refs.\ci{mueller,radyushkin99,musatov} for the motivation
to this approach. The GPDs constructed this way satisfy the polynomiality requirement,
i.e. $x$-moments of the GPDs are polynomials in $\xi$. For $H$ and $E$ there is an 
additional D-term~\cite{polyakov-weiss} which lives only in the region
$-\xi< x <\xi$ as implied by the Heaviside function $\Theta$. It provides a term 
$\sim \xi^{n+1}$ (for even $n$) to the $n$-th moment of these GPDs,
while the integral term in Eq.~\req{eq:theory-DD} provides a polynomial 
in $\xi$ only of order $n$. For (quarks) gluons the $D$-term is (anti)symmetric
in the argument $x/\xi$ and contributes with opposite sign to $H$ and $E$. There is no
$D$-term for the other GPDs. For valence quarks there is no  $D$-term at all. According to
the definition of the gluon GPDs there is an extra factor $|\xi|$ in front of the gluonic $D$-term.

For the weight function, $w$, that generates the skewness dependence, the form
\be
w_(\rho,\eta)\=\frac{\Gamma(2n+2)}{2^{2n+1}\Gamma^2(n+1)}\,
             \frac{[(1-|\rho|)^2-\eta^2]^{n}}{(1-|\rho|)^{2n+1}}
\ee
is often used, e.g. \ci{GK1,GK3,VGG98,gui-rad}. The power $n$ is regarded as a free 
parameter.

\begin{figure}[htb]
\begin{center}
\includegraphics[width=0.36\tw]{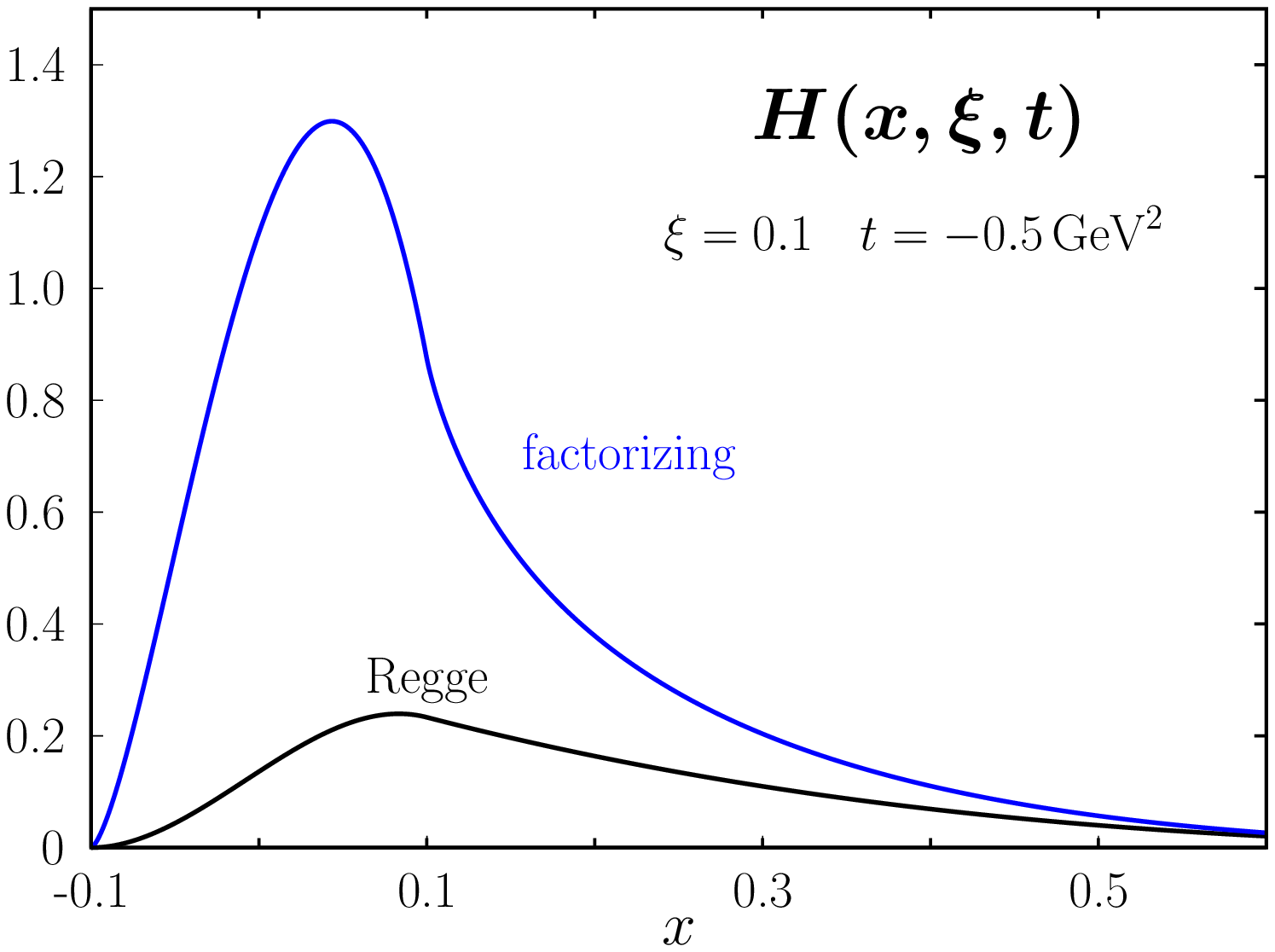}\hspace*{0.05\tw}
\includegraphics[width=0.35\tw]{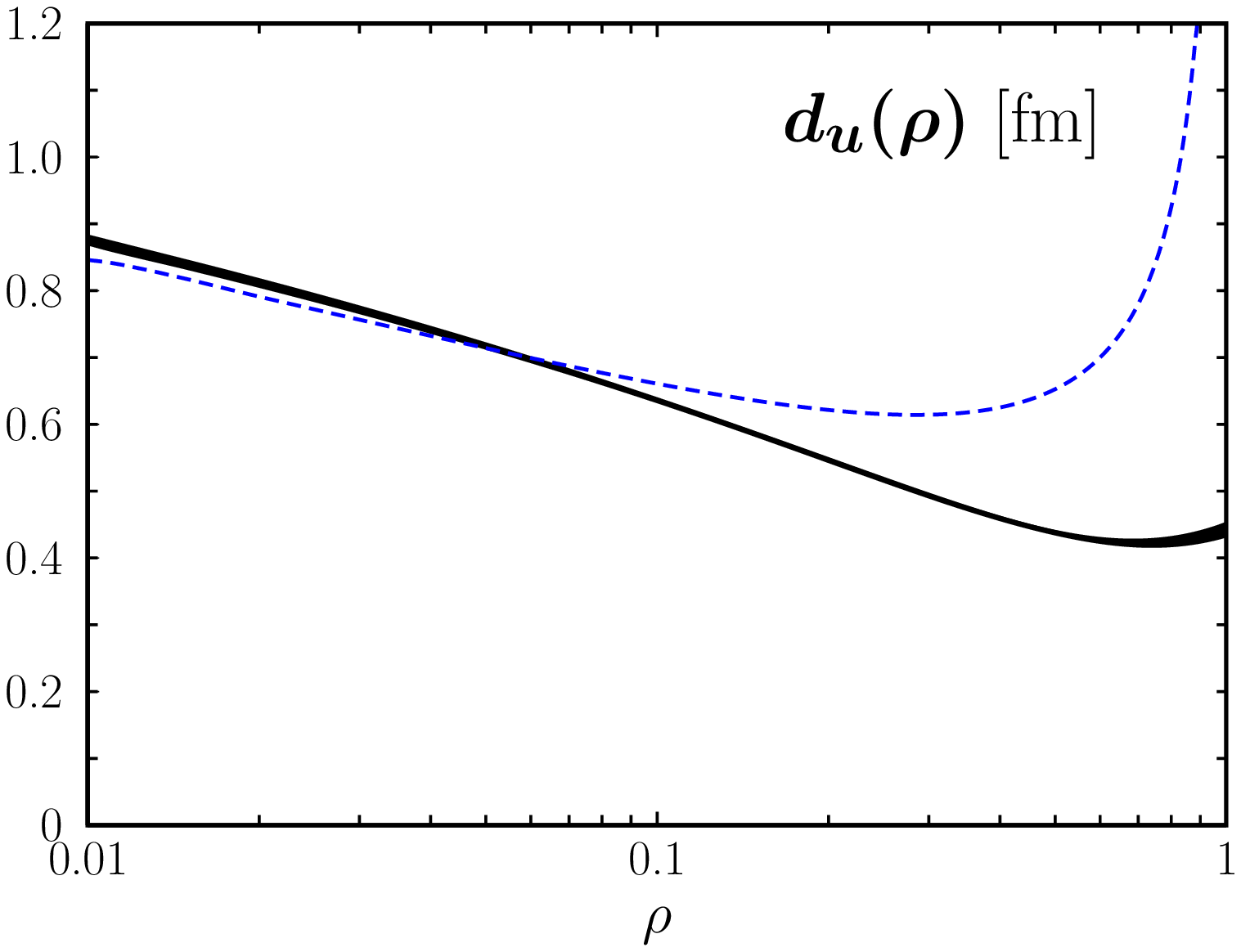}
\end{center}
\vskip -0.03\tw
\caption{Left: A valence-type GPD constructed from the $\rho$-$t$ factorizing and
from the Regge-type double distribution at $\xi=0.1$ and $t=-0.5\,\gev^2$. The Regge
trajectory is assumed to be $\alpha=0.5+t$ GeV$^{-2}$. Right: The average distance
$d_u$ for $u$-valence quarks. The solid line is the profile function of 
Eq.~\req{eq:theory-dfjk-profile} with $A_u=1.264\,\gev^{-2}$, $B_u=0.545\,\gev^{-2}$, 
$\alpha'=0.961\,\gev^{-2}$; the dashed line was obtained from Eq.~\req{eq:theory-regge-profile} 
with $B=0$, $\alpha'=0.9\,\gev^{-2}$.}  
\label{fig:theory-geometry} 
\end{figure}

For the $t$-dependence, as the most simple ansatz for a zero-skewness GPD, one 
originally considered the product of 
its forward limit and a $t$-dependent form factor, e.g.\ $H^q(x,\xi=0,t)=q(\rho)F_1^{q}(t)$, 
\ci{VGG98}-\ci{kugler}. Such an ansatz, factorizing in $\rho$ and $t$, is 
evidently in conflict with Burkhardt's observation of a $\rho-t$ correlation
\ci{burkhardt,burkhardt02}. In later work a more complicated ansatz has been exploited
\be
K(\rho,\xi=0,t) = k(\rho) \exp{[f(\rho)t]}
\label{eq:theory-ansatz}
\ee
where $k(\rho)$ is either the appropriate parton distribution function (PDF) for $H$, 
$\widetilde H$ and $H_T$ or, for 
the other GPDs, parametrized like the PDFs with some parameters to be adjusted to data:
$N\rho^{-\alpha(0)}(1-\rho)^{\beta}$ (for gluon GPDs an extra power of $\rho$ appears). 
The profile function, $f(\rho)$, is frequently parametrized 
in a Regge-like manner, e.g.\ \ci{GK1,GK3,gui-rad,goeke01},                 
\be
f(\rho)\=B+\alpha^\prime\ln{(1/\rho)}\,.
\label{eq:theory-regge-profile}
\ee
Since the function $k$ behaves as $\sim \rho^{-\alpha(0)}$ at small $\rho$
we have altogether a behavior as $\sim \rho^{-\alpha(t)}$ for the zero-skewness GPD with 
a linear Regge trajectory. Consider for example a typical Regge trajectory, 
$\alpha\simeq 0.5 + t$ GeV$^{-2}$. In this case the zero-skewness GPD behaves as $\sim 1/\sqrt{\rho}$
for $t\to 0$. For increasing $-t$ the singularity at $\rho=0$ becomes milder and vanishes
for $t\simeq -0.5\gev^2$. This feature is to be contrasted with the results from the
$\rho-t$ factorizing ansatz for which the zero-skewness GPD behaves as $\sim 1/\sqrt{\rho}$ for 
all $t$ with strong consequences for the resulting GPD. This can be seen from Fig.\ 
\ref{fig:theory-geometry}
where the GPDs from the factorizing and from the Regge-like ansatz are compared to each
other at $\xi=0.1$ and $t=-0.5\,\gev^2$. In both cases the same forward limit is used,
i.e.\ the GPDs fall together at $t=0$. With increasing $-t$ the GPDs differ 
in magnitude from each other more and more. 

Although the Regge-like profile function possesses a $\rho - t$ correlation 
- small (large) $-t$ go along with small (large) $\rho$, it still has deficiencies at large 
$\rho$. To see this, we consider the GPD $H$ for a quark flavor $q$ 
as an example. The average impact parameter of $q(\rho,\vbs)$,
the Fourier transform of $H^q$ with respect to the momentum transfer,
at a given $\rho$, is~\cite{burkhardt02} 
\be
\langle b^2\rangle_\rho\=\frac{\int d^2\vbs \vbs^2 q(\rho,\vbs)}
                               {\int d^2\vbs q(\rho,\vbs)} \=4f(\rho) 
\label{eq:impact}
\ee
In the proton's center of momentum the relative distance between the struck quark
and the spectator system is $\vbs/(1-\rho)$. Its average
\be
d(\rho)\=\frac{\sqrt{\langle \vbs^2 \rangle_\rho}}{1-\rho}
\label{eq:drho}
\ee
can be regarded as an estimate of the size of the proton. For the profile function
of Eq.~\req{eq:theory-regge-profile} the distance $d_q$ is singular for $\rho\to 1$.
Because of this unphysical behavior the Regge-like ansatz cannot be used in the large 
$\rho$ region.

In order to cure this problem the following profile function has been advocated for
in Refs.~\ci{DFJK4,DK13} and utilized in an GPD analysis of the nucleon form factors:
\be
f(\rho)\=(B+\alpha^\prime\ln{(1/\rho)})(1-\rho)^3 + A\rho (1-\rho)^2
\label{eq:theory-dfjk-profile}
\ee  
For this profile function the average distance between the struck quark and the 
cluster of spectators tends to the constant $A$ for $\rho\to 1$, see Fig.\ 
\ref{fig:theory-geometry}. The first term of Eq.~\req{eq:theory-dfjk-profile} dominates at
low $\rho$ and approximately falls together with the Regge-like profile function
while at large $\rho$ the second term becomes dominant. In Ref.~\ci{selyugin} a profile 
function with just a $(1-\rho)^2$ term is discussed.  
The nucleon form factors are important constraints of the valence quark GPDs 
$H, E, \widetilde H$. The sum rules for the form factors allow to determine these GPDs
at zero skewness for a given parametrization. For detailed analyses, see refs.~\ci{gui-rad,DFJK4,DK13}. 
Used as input to the double distribution representation of Eq.~\req{eq:theory-DD} the full valence
quark GPDs can be evaluated and applied in analyses of hard exclusive reactions. 
Its application in deeply virtual meson production is discussed in the following sections.

An alternative parameterization of the GPDs starts from their double partial wave 
expansion in the conformal and in the $t$-channel SO(3) partial waves 
expansion \ci{mueller:14}. However, this parametrization has been rarely used
in actual analyses of DVMP. 
It goes without saying that all these ans\"atze hold at a low scale, typically 
a few GeV. The GPDs at larger scales are to be evaluated from the evolution equations.
For instance, this can be achieved by using the Vinnikov code \ci{vinnikov}.


\section{Review of Vector Mesons data and interpretation}

\subsection{Integrated cross section $\sigma$}

\begin{figure}[!h]
\begin{center}
\includegraphics[width=0.42\tw]{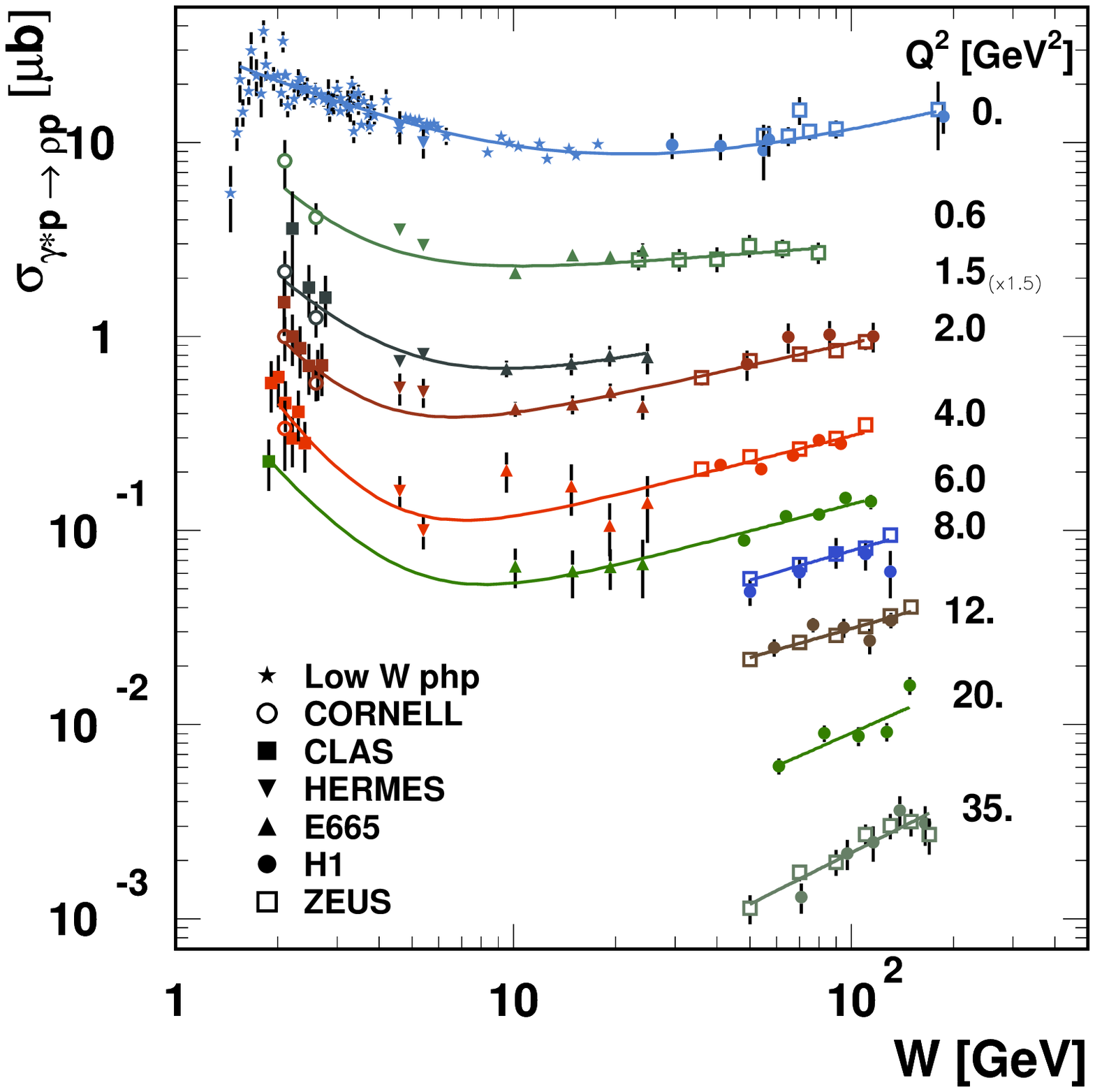}
\includegraphics[width=0.42\tw]{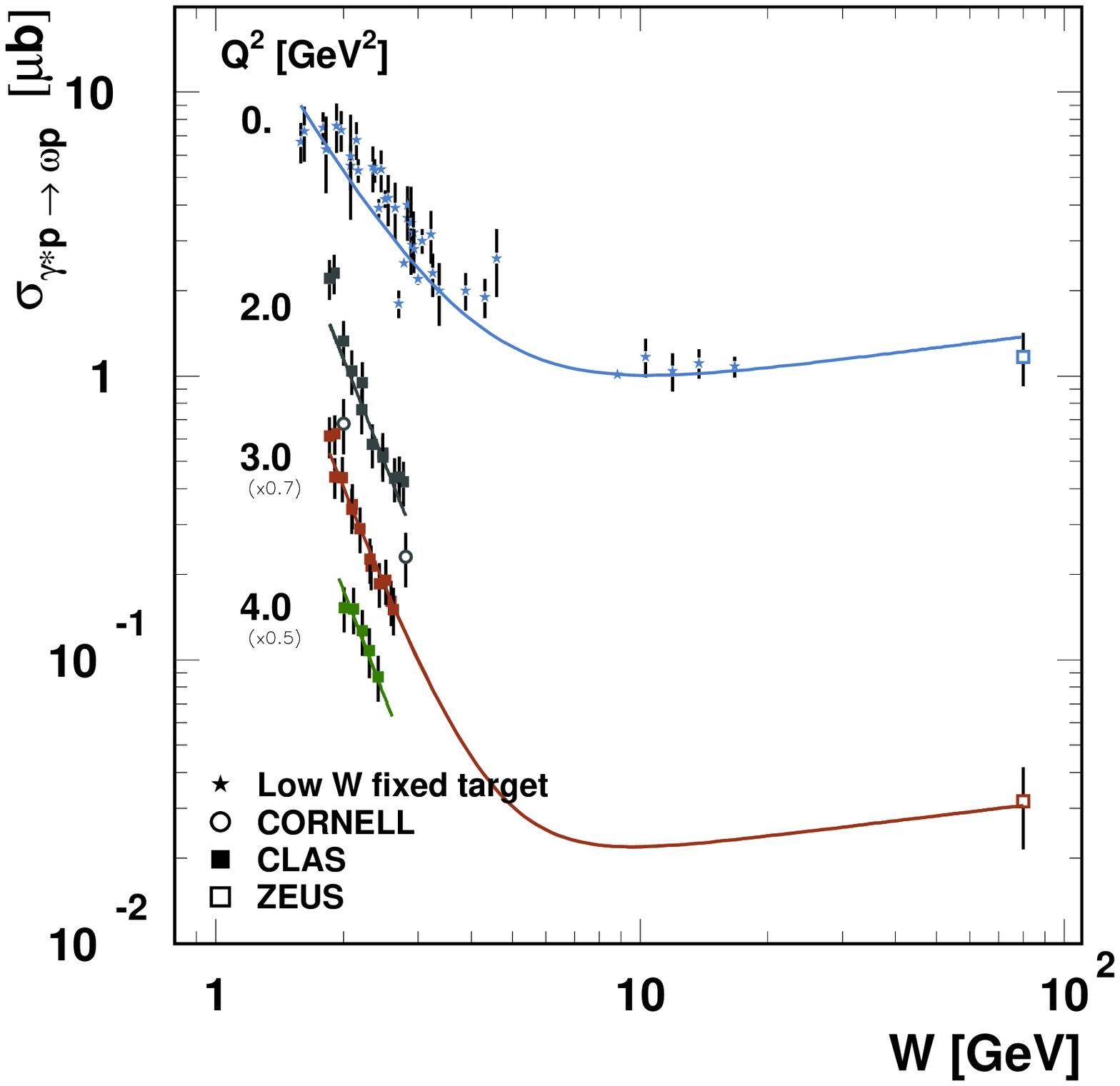}
\includegraphics[width=0.42\tw]{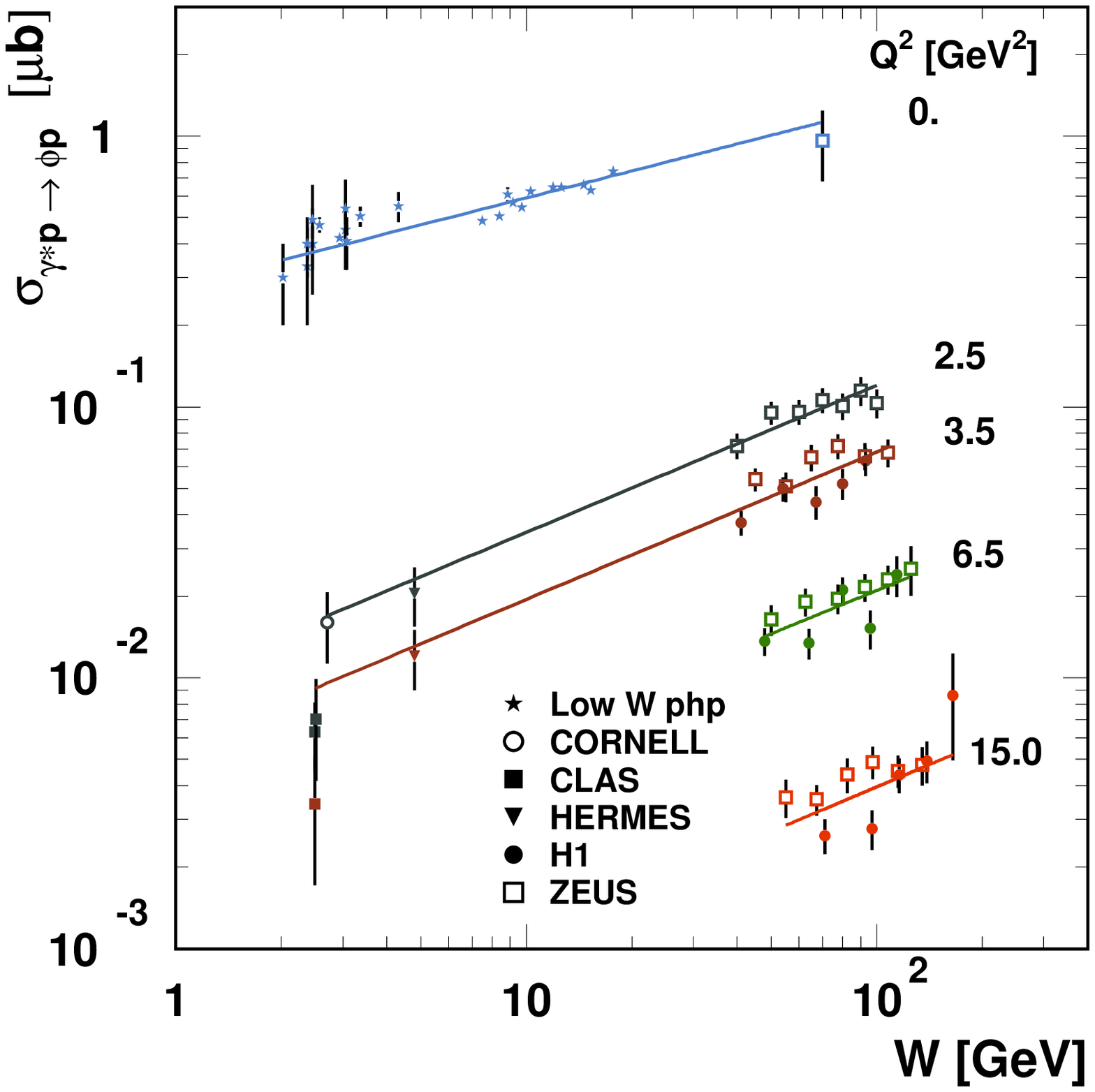}
\includegraphics[width=0.42\tw]{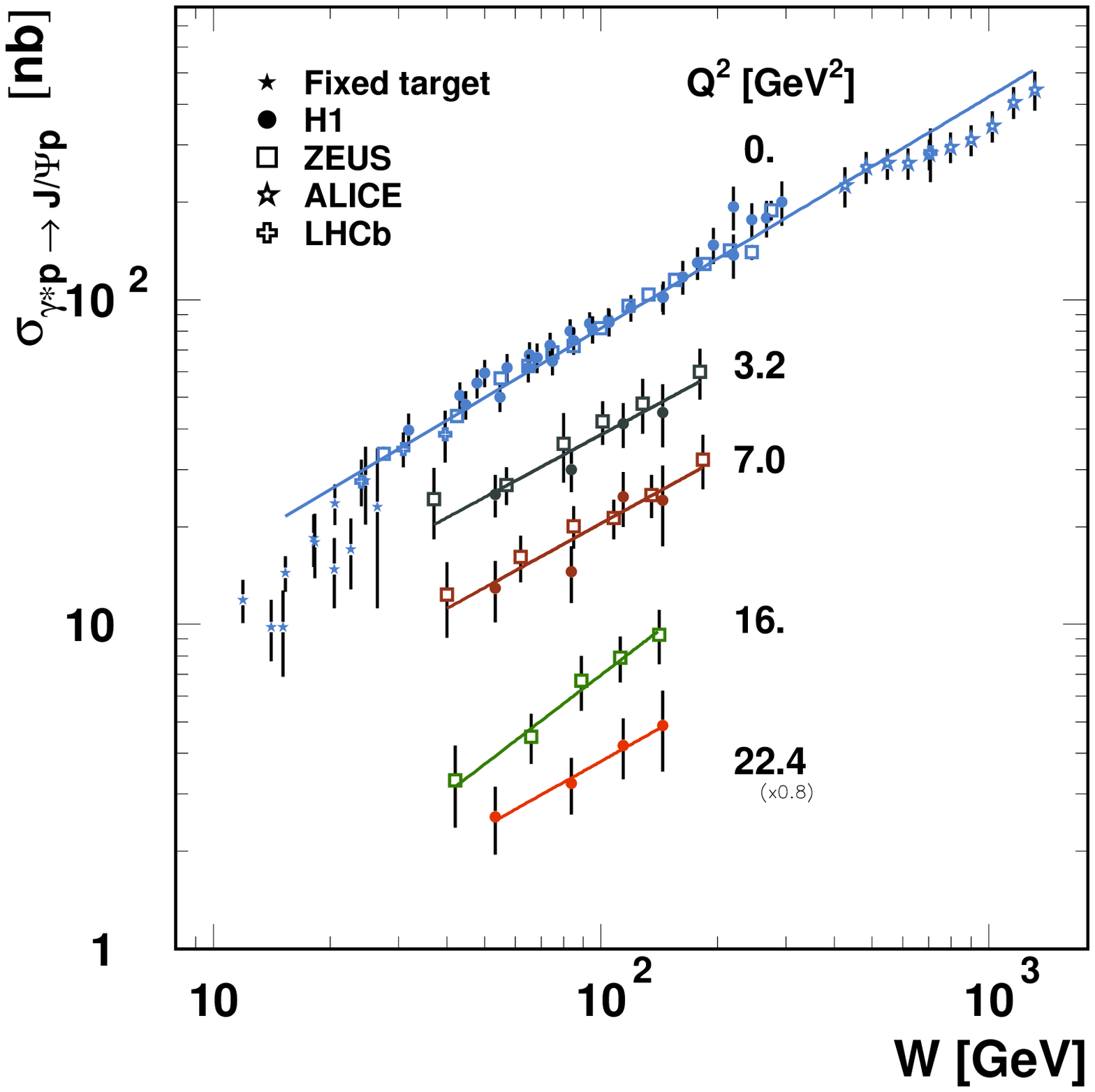}
\caption{Top left: $W$-dependence of the $\gamma^{(*)}p\to p\rho^0$ cross sections
for different $Q^2$ values (``php" denotes photoproduction).
Some cross sections have been rescaled in $Q^2$ in order to
compare between different experiments at the same $Q^2$. See text for data references.
Top right: same for the $\omega$ channel. Bottom left: same for the $\phi$ channel.
Bottom right: same for the $J/\Psi$ channel. 
The curves indicate the results of the fit to the function of Eq.~\req{eq:wfit}.
The fits are done for $W$ values large enough to be insensitive to threshold effects.}
\label{fig:vmwdep} 
\end{center}
\end{figure}

\begin{figure}[!h]
\begin{center}
\includegraphics[width=0.55\tw]{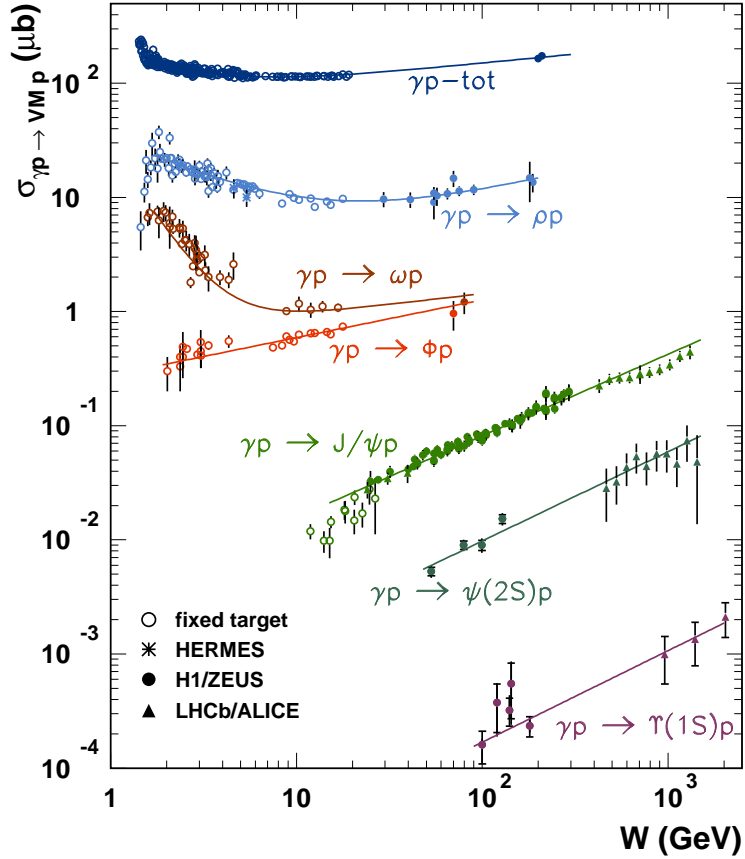}
\caption{$W$-dependence of the photoproduction cross sections 
for the reactions $\gamma p\to p\rho^0, \omega, \phi,
J/\Psi, \Psi(2S), \Upsilon$. The total $\gamma p$ cross section is also
shown. The curves indicate the results of the fit to 
the function of Eq.~\req{eq:wfit}.}
\label{fig:photo} 
\end{center}
\end{figure}

Due to a relatively high cross section, 
exclusive lepto- and photoproduction of vector meson 
on the proton is one of the exclusive processes the most 
studied experimentally. 
The cross sections for $\rho^0$, $\omega$, $\phi$, 
$J/\Psi$, $\Psi(2S)$, $\Upsilon$ production have been measured 
over a wide range of energies, from threshold 
up to $W\approx$ 200 GeV for light mesons and, taking into account 
the recent LHC data, up to about 1 TeV for the heavy mesons.
For light vector mesons, $Q^2$ ranges from $Q^2$=0 GeV$^2$ up 
to 100 GeV$^2$ in the high $W$ domain, where there are data from 
the H1~\ci{h1a,h1b,h1c}, ZEUS~\ci{zeusa,zeusb,zeusc,zeusomea,zeusomeb,zeusphia,zeusphib} 
and HERMES\ci{HERMESrho} (HERA) and E665 (Fermilab)\ci{e665} 
experiments. At low $W$, experiments at CORNELL~\ci{Cassel:1981sx} 
and with the CLAS detector at Jefferson Lab~\cite{rhoclas1,Morrow:2008ek,omegaclas,phiclas1,phiclas2}  
have measured the $\rho^0$, $\omega$ and $\phi$ electroproduction 
channels up to $Q^2$=5 GeV$^2$. 
For the heavy mesons, there are electroproduction data only for
the $J/\Psi$ from HERA~\cite{jpsih1a,jpsizeusb}; otherwise, there are photoproduction 
data coming from the H1~\cite{jpsih1b}, ZEUS~\cite{jpsizeusa}, 
LHCb~\cite{Aaij:2013jxj,Aaij:2014iea,Aaij:2015kea,McNulty:2015cpd} 
and ALICE~\cite{TheALICE:2014dwa} experiments and a few
fixed target data at low energies.

Fig.~\ref{fig:vmwdep} shows representative data in leptoproduction, with
some photoproduction data as well, and
Fig.~\ref{fig:photo} shows all photoproduction data.
For the $\rho^0$ and $\omega$ channels,
two regimes are clearly apparent, in both lepto- and photoproduction. 
Starting from threshold, after
a rapid rise due to the opening of the phase space, the cross sections 
decrease from $W\approx$ 2 GeV down to $W\approx$ 7 GeV. Then, the 
cross sections slowly rise with energy. 
For the other vector meson channels, $\phi$, $J/\Psi$, $\Psi(2S)$ and $\Upsilon$,
above the threshold effect,
there is only one behavior of the cross section: a steady rise with $W$
from threshold up to the highest energies measured. 
One can clearly notice that the slope of the $W$-dependence
at large $W$ increases with $Q^2$ for the $\rho^0$ electroproduction channel
and with the mass of the vector mesons for the photoproduction data.
This indicates that the mass of the heavy mesons acts,
like $Q^2$, as a hard scale.

\begin{figure}[h!]
\begin{center}
\epsfxsize=10. cm
\epsfysize=10. cm
\centerline{\epsffile{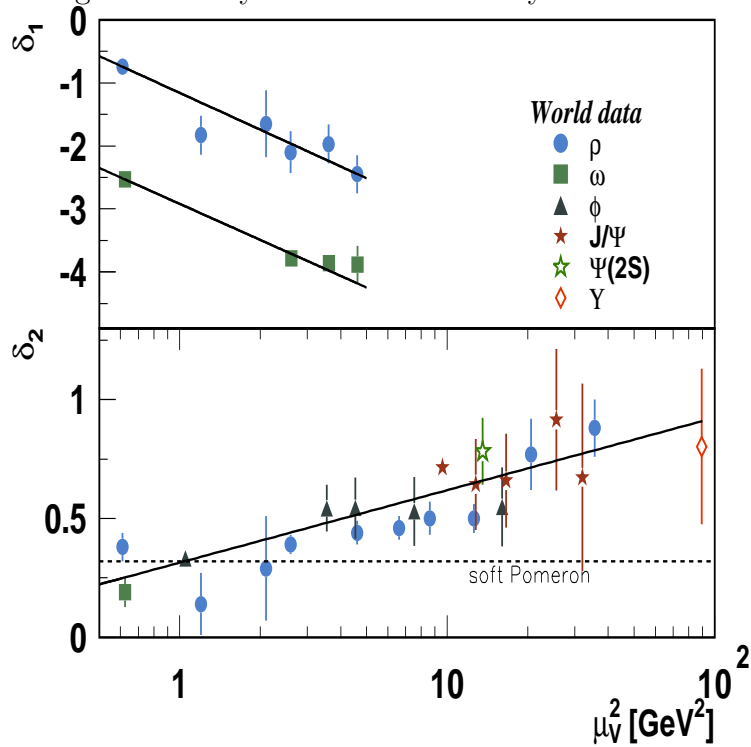}}
\caption{The $Q^2$-dependence of the powers $\delta_1$ and $\delta_2$ 
reflecting the $W$-dependence of the exclusive vector meson cross sections.
The solid lines show fits of the form: $\delta(\mu^2_V)=A + B \ln(\mu^2_V)$ ($\mu^2_V$ in GeV$^2$).
For $\delta_2$, one finds $A=0.31\pm 0.02$ and $B=0.13\pm 0.01$.} 
\label{fig:delta} 
\end{center}
\end{figure}

In order to be quantitative, the cross section data in Figs.~\ref{fig:vmwdep} 
and~\ref{fig:photo} have been fitted with the following function:

\be
 \sigma_V(W,\mu^2_V)= a_1 W^{\delta_1(\mu^2_V)} + a_2 W^{\delta_2(\mu^2_V)}
\label{eq:wfit}
\ee

with $a_1$, $a_2$, $\delta_1$ and $\delta_2$ as free parameters.
We define the scale $\mu_V^2$ as $Q^2+M_V^2$,
where $M_V$ is the mass of the vector meson, in correspondance
with what we observed in Figs.~\ref{fig:vmwdep} and~\ref{fig:photo}.
The light-meson mass is regarded as representative for the
hadronic scale at which meson photoproduction occurs.
In order to reduce the model dependence in the LHCb measured cross section as a
function of $W$ for the $J/\Psi$ and $\Psi(2S)$~\cite{Aaij:2013jxj,Aaij:2014iea,McNulty:2015cpd} 
to a negligible level with respect to the other systematics, 
only the high $W$ solution discussed in Refs.~\cite{Aaij:2013jxj,Aaij:2014iea,McNulty:2015cpd} 
is kept in the fit.
The curves in Fig.~\ref{fig:vmwdep} and ~\ref{fig:photo} show the results of these fits.
Fig.~\ref{fig:delta} shows the dependence of $\delta_1$ and $\delta_2$ 
on $\mu_V^2$. For the $\phi$ and the heavy mesons the parameter $a_1$ is zero.
The power $\delta_1$, characterizing
    the low energy behavior of the $\rho^0$ and $\omega$ cross sections,
    differs for these two reactions and increases in absolute value with 
    increasing $Q^2$. To date, there is no theoretical explanation for $\delta_1$.
    In contrast to the behavior of $\delta_1$, the power $\delta_2$
    is approximately the same for the 
    $\rho^0$ and $\omega$ cross sections as well as 
    for the $\phi$ and heavy meson
    cross sections. This universal behavior of $\delta_2$ will
    be explained below. It is, however, interesting 
    to note that a recent article 
(Jones et al.~\cite{Jones:2015nna})
analysed the $J/\Psi$ and $\Upsilon$ photoproduction data including
those from the LHC~\cite{Aaij:2013jxj,Aaij:2014iea,Aaij:2015kea,TheALICE:2014dwa}. The handbag
amplitudes have been calculated to NLO accuracy~\cite{Ivanov:2004vd}. It seems
that the predictions for the $J/\Psi$ cross sections underestimate the data,
while good agreement with the $\Upsilon$ is found.
According to~\cite{Jones:2015nna}, this indicates that the gluon density in recent
PDF analysis is too small at low scales.

\subsection{The longitudinal and transversal cross sections $\sigma_L$ and $\sigma_T$}

\begin{figure}[htb]
\begin{center}
\includegraphics[width=0.5\tw]{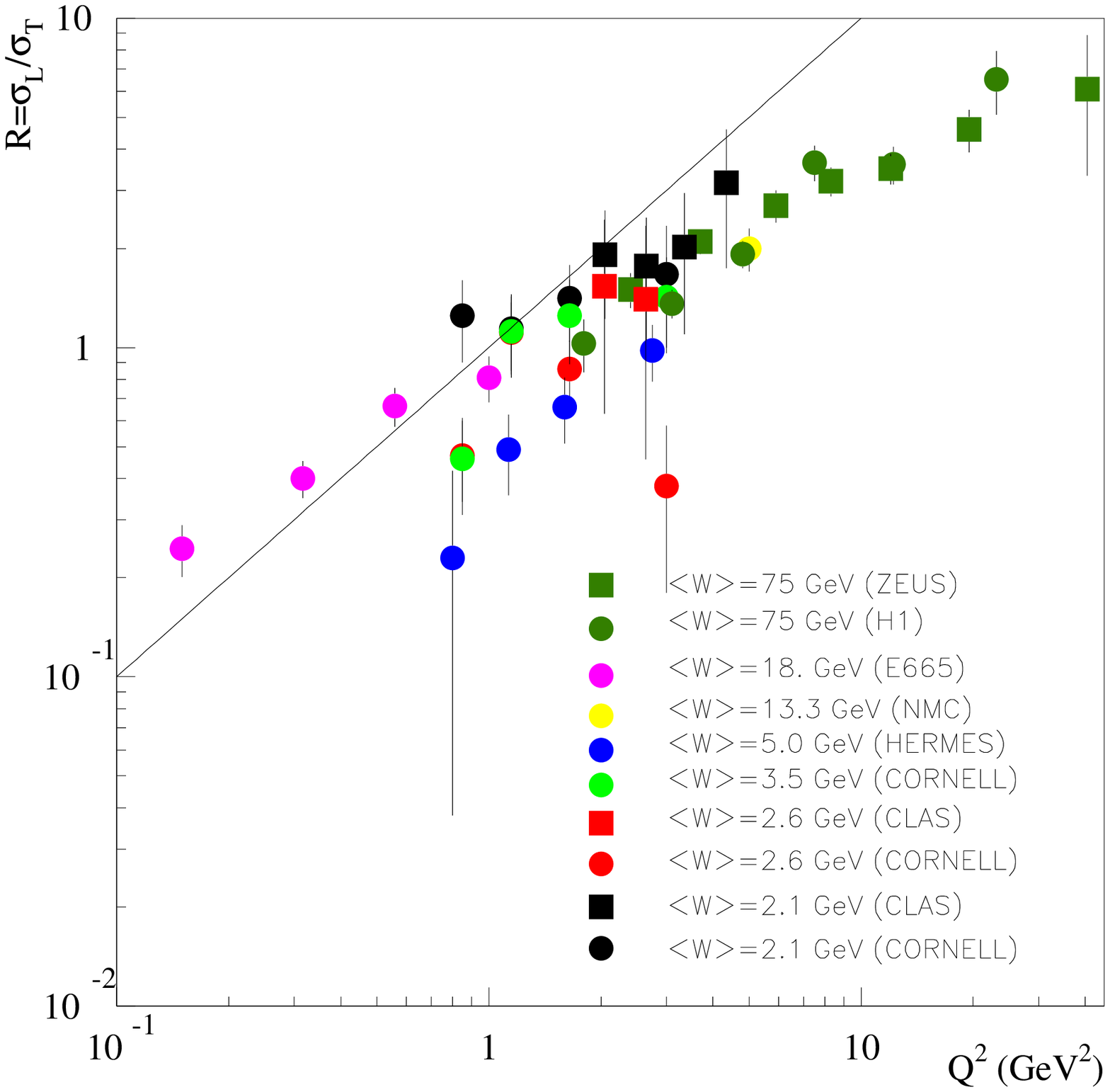}
\includegraphics[width=0.45\tw]{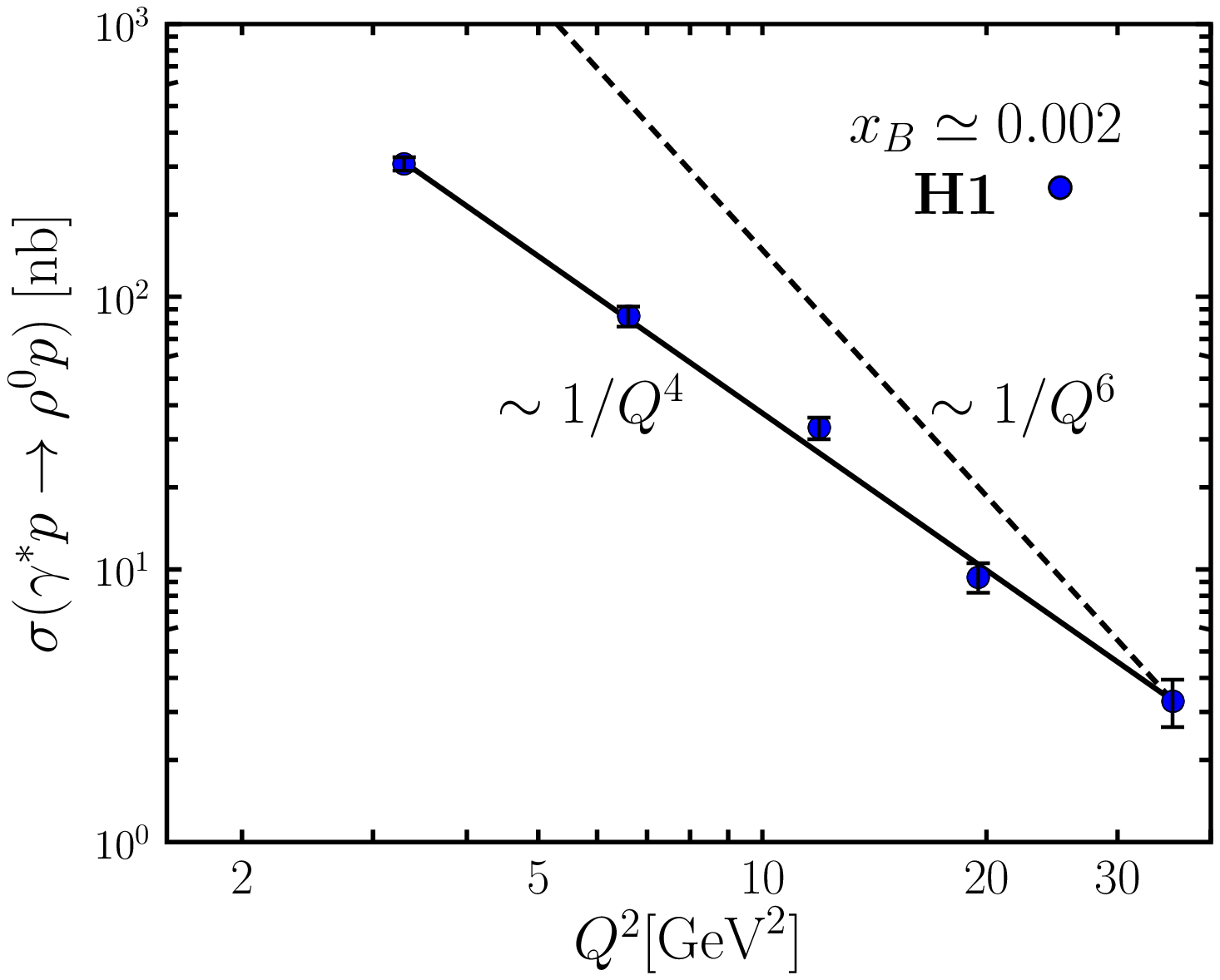}
\end{center}
\vskip -0.03\tw
\caption{Left: The ratio $\sigma_L/\sigma_T$ for $\rho^0$ leptoproduction
versus $Q^2$. The solid line represents a $\propto Q^2$ dependence
with arbitrary normalisation. Right: Measured $Q^2$-dependence of the $\rho^0$ cross section 
at fixed $x_B\approx$0.002. Data are from Refs.~\cite{rhoclas1,Morrow:2008ek} (CLAS),
~\cite{Cassel:1981sx} (CORNELL),~\cite{HERMESrho} (HERMES),~\cite{nmc} (NMC),
~\cite{e665} (E665),~\cite{h1a,h1c} (H1) and~\cite{zeusc} (ZEUS).} 
\label{fig:siglovert} 
\end{figure}

A feature to look at, in order to test the applicability 
of the handbag formalism, is whether the predicted dominance of the longitudinal part of the cross section can be observed experimentally.
Most of the $\rho^0$ data of Fig.~\ref{fig:vmwdep} have been L/T separated. 
Indeed, the analysis of the decay angular distribution of the vector mesons
gives access to the polarization states of the vector mesons. Then, assuming
that the helicity between the final state vector meson and the initial virtual 
photon is conserved, one can deduce the polarization of the virtual
photon, and thus $\sigma_L$ and $\sigma_T$. This property is referred to as $s$-channel 
helicity conservation (SCHC). It can be checked experimentally by looking
at specific decay matrix elements of the vector mesons~\cite{Schillel}
which are sensitive to 
$\gamma^{*}_L\to V_T$, $\gamma^{\*}_T\to V_L$ or 
$\gamma^{*}_T\to V_{-T}$ transitions. SCHC has 
been found to hold sufficiently well for $\rho^0$ and $\phi$
production to allow for the L/T separation of the cross sections.
Fig.~\ref{fig:siglovert} (left) presents the $Q^2$-dependence of 
the ratio $\sigma_L/\sigma_T$ for $\rho^0$ production. 
It is predicted that this ratio should grow $\propto Q^2$
both as $Q^2\to 0$ and, as predicted by the handbag approach$, Q^2\to \infty$. 
This is realized at low $Q^2$ but the ratio is much milder than a $\propto Q^2$
behavior for $Q^2\to\infty$.
Though, taking into account parton transverse momenta, the handbag 
calculations\cite{GK3,martin-ryskin-teubner} can be put in agreement with the data. 
The cross section $\sigma_L$ 
is larger than $\sigma_T$ for $Q^2\approx$1.5 GeV$^2$.
The transverse cross section $\sigma_T$ remains sizeable up to $Q^2\approx$ 40 GeV$^2$,
the maximum $Q^2$ value at which it has been measured.

Another interesting feature to look at is the $Q^2$-dependence of the longitudinal cross section data.
We recall that a prediction of the handbag approach is that the longitudinal cross section
should scale, at fixed $x_B$, as $1/Q^6$ (up to evolution effects).
In Fig.~\ref{fig:siglovert} (right), the H1 data for the $\rho^0$ cross section are shown at 
 fixed $x_B\simeq$ 0.002. The data span more than an order of magnitude in $Q^2$. One 
  sees that the cross section falls approximately as $1/Q^4$, i.e. the $Q^2$ slope
  is milder than what is predicted by the leading-twist handbag formalism. 
Given the behavior of $\sigma_L/\sigma_T$, i.e. the $\sigma_T$ contribution
is all the more important as $Q^2\to$ 0, the experimental $Q^2$-dependence of $\sigma_L$
is even flatter than $1/Q^4$.
One may wonder whether higher-order perturbative corrections are responsible for this effect.
As shown by Diehl and Kugler \ci{diehl-kugler}, due to large BFKL-type logarithms, the NLO 
corrections to  $\sigma_L$ are huge at small skewness with opposite sign as
compared to the LO term. 
A comparison with experiment however requires a resummation of these logarithms which has 
not yet been performed. An alternative concept is advocated for in \ci{Meskauskas:2011aa}. 
In this work the GPD $H$, 
parametrized according to the double partial wave expansion and fitted to the HERA data on 
DVMP and DVCS to leading-twist accuracy, exhibit strong evolution effects. 
This means that the reduction 
of the $1/Q^6$ fall to an effective $1/Q^4$ one is realized by logarithms of $Q^2$. 
It remains to be seen whether this 
concept can be extended to smaller $W$. It should also be mentioned that a similar fit to only the 
DVCS data leads to different GPDs with much milder evolution effects.

\begin{figure}[h!]
\begin{center}
\includegraphics[width=0.36\tw]{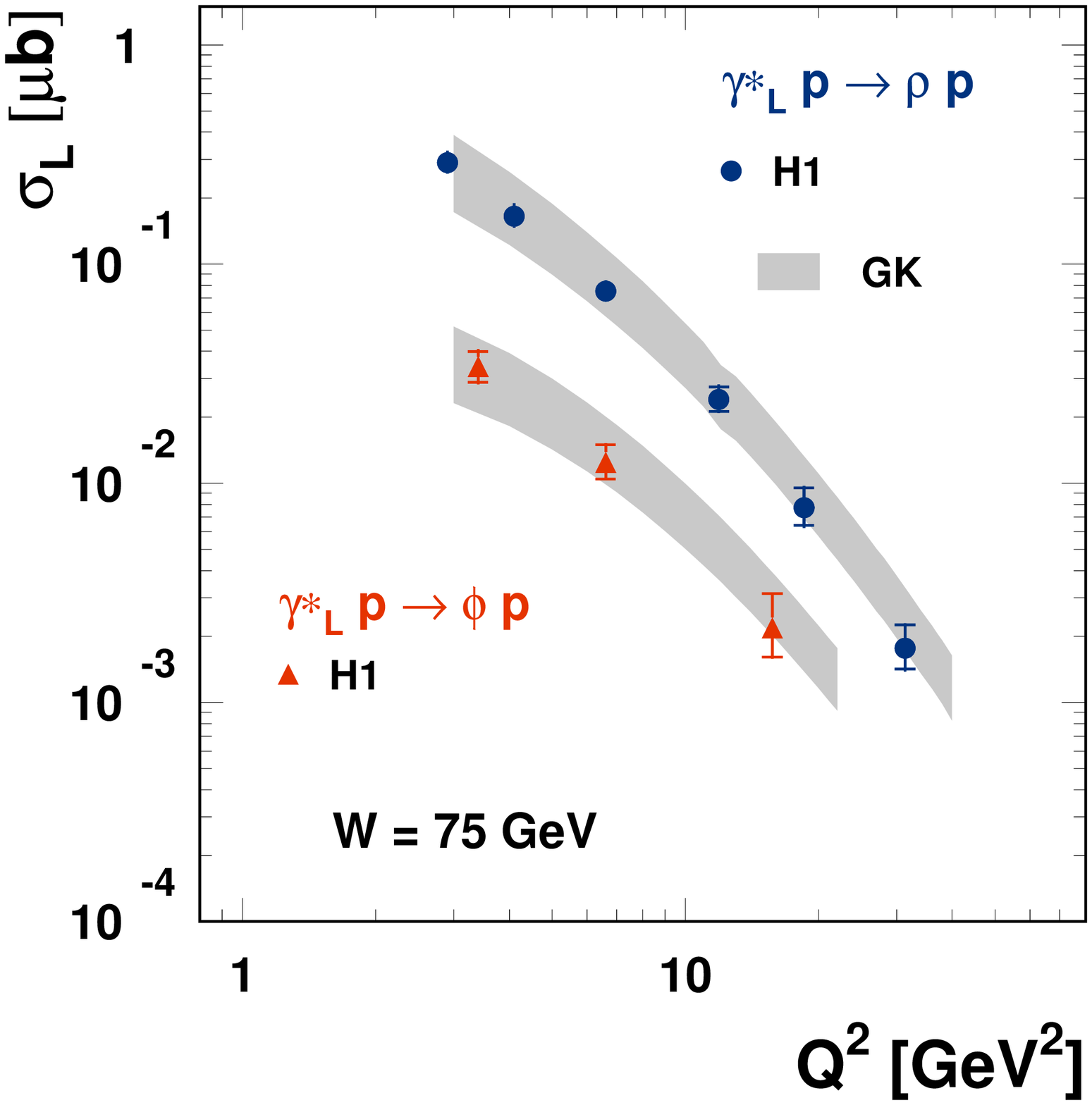}
\includegraphics[width=0.36\tw]{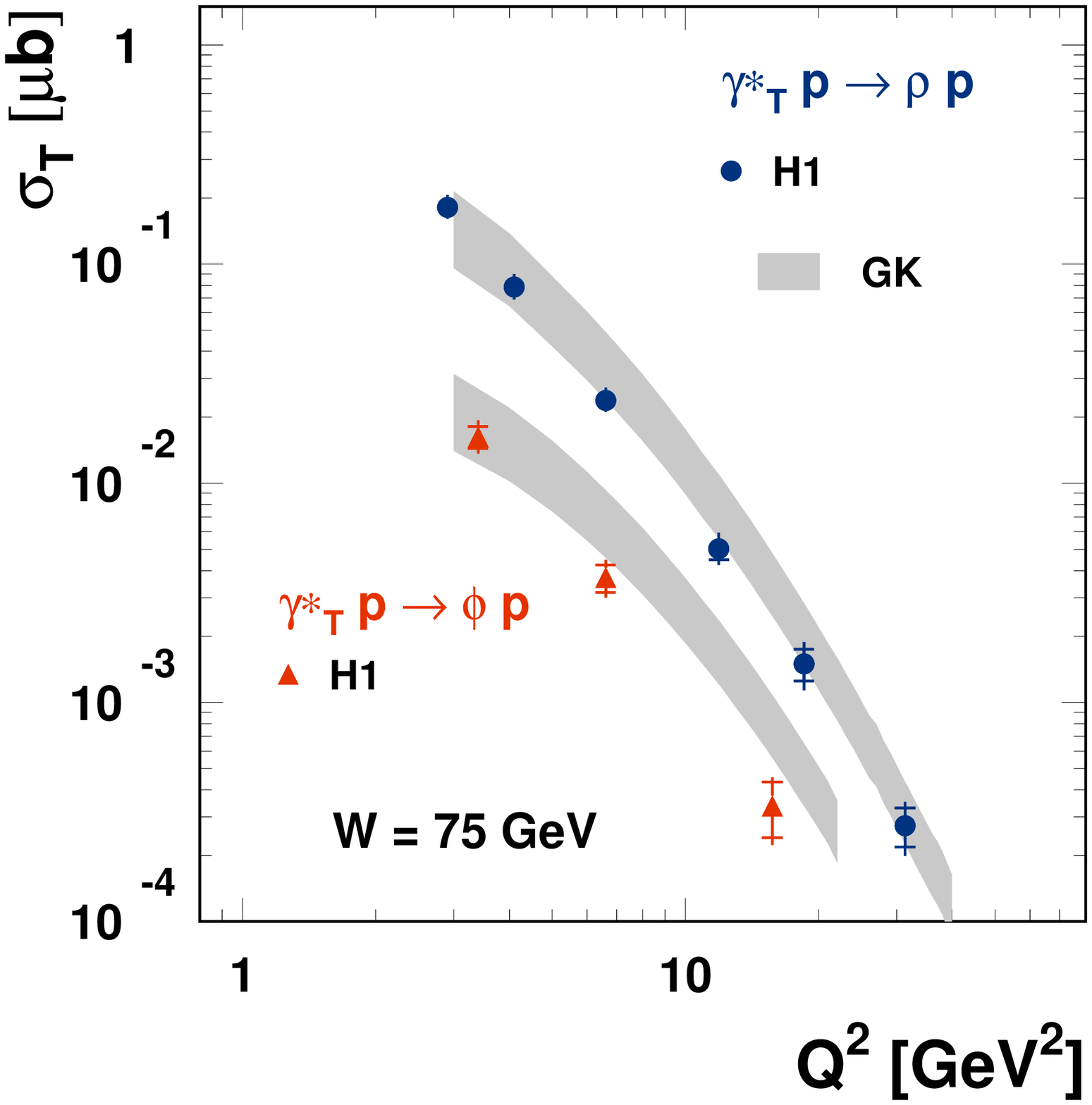}
\caption{Left (right): comparison of the $Q^2$-dependence at fixed $W$=75 GeV 
with the GK model for $\sigma_L$ ($\sigma_T$) of the $\rho^0$ and $\phi$ cross sections.
The grey bands indicate the parametric uncertainties  of the GK results~\ci{GK3}.
Data are from Ref.~\cite{h1c} (H1).}
\label{fig:q2} 
\end{center}
\end{figure}
\begin{figure}[h!]
\begin{center}
\epsfxsize=8. cm
\epsfysize=8. cm
\centerline{\epsffile{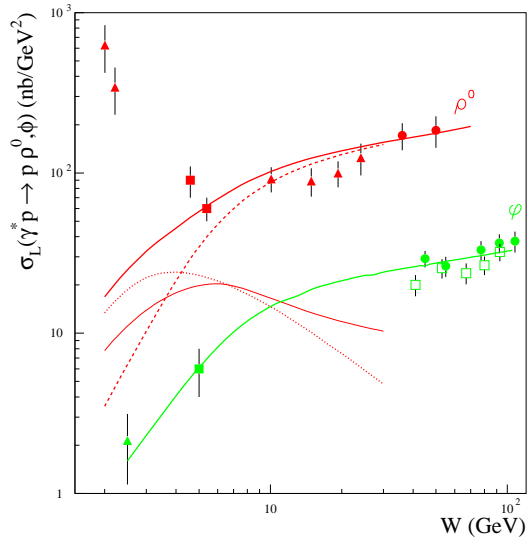}}
\caption{$W$-dependence at $Q^2\approx$3.8 GeV$^2$ of the $\gamma^*p\to p (\rho^0, \phi)$ cross sections 
in (red, green) compared to GK~\cite{GK3,GK2} (thick solid curves) and VGG~\cite{VGG98,gui-rad} 
(red thin solid curve) calculations (the VGG calculation contains only quark exchange).
The dashed (dotted) line represents the gluon plus sea (valence-quark
plus (gluon + sea)-valence interference) contribution.}
\label{fig:vgggk} 
\end{center}
\end{figure}

In \ci{GK2} it is suggested that the $\approx 1/Q^4$ behavior is generated by 
meson size effects modeled by quark transverse momentum in the subprocess while the emission 
and reabsorption of the partons from the proton are still treated in collinear approximation. 
The quark transverse momenta in the subprocess result in a separation of color sources in
the impact parameter plane  which is accompanied by gluon radiation. This is taken into
account in \ci{GK2} by a Sudakov form factor which includes the gluon radiation resummed to all
orders of perturbation theory in next-to-leading-log approximation. Instead of distribution
amplitudes $\vk$-dependent light-cone wave functions have to be used with a parameter describing
the transvere size of the meson. This approach 
turns into the leading-twist handbag result asymptotically. Its gluonic part
bears resemblance to the color-dipole model 
\ci{nemchik-nikolaev-predazzi-zakharov,Strikman:2009bd}. The approch 
has been generalized~\ci{GK3} in order to account also for
     transversally polarized photons. The infrared singularity appearing in 
     collinear approximation in the
   corresponding amplitudes are regularized by the parton transverse momenta.
   The generalized handbag approach succesfully describes all $\phi$ and $\rho^0$
   leptoproduction data, cross sections, spin density matrix elements
   and spin asymmetries for $W\geq$ 2 GeV and $\geq$ 4 GeV, respectively. As an
   example, we show in Fig.~\ref{fig:q2} results from Ref.~\ci{GK3} and compare them to the
   HERA data. It is seen that it reproduces
very well, in addition to the normalisation, the $Q^2$-dependence 
of $\sigma_L$ and $\sigma_T$. In Fig.~\ref{fig:vgggk}, we show the $W$-dependence 
of the longitudinal cross section 
for the $\rho^0$ and $\phi$ channels compared to this approach.
There is a good agreement over the whole $W$-range for the $\phi$-channel
(for which there is only gluons exchange) 
but only for $W\gsim$ 4 GeV for the $\rho^0$ channel. 

At large energies and small $x_B$, the $W$-dependence of the cross sections 
shown in Figs.~\ref{fig:vmwdep}, ~\ref{fig:photo} and~\ref{fig:vgggk}
can be understood as follows: vector meson
   production in this kinematical region is a diffractive process with
   a dominantly imaginary amplitude which is fed by the gluonic GPD $H^g$.
   As can be seen from Eqs.\req{eq:amplitudes} and~\req{eq:leading-twist},
    the imaginary part of the $\gamma^*p\to Vp$
   amplitude is proportional to $H^g$ for $x=\xi$:

\be
Im {\cal M}^V_{0+,0+}\sim H^g(\xi,\xi,t)/\xi
\ee
From Eq.\req{eq:theory-DD} one derives the small $\xi$ behavior of the GPDs~\ci{kroll14}
\be
H^g(\xi,\xi,t)\=c_g(\delta_g,\alpha'_g,n_g)2\xi g(2\xi)e^{t(B-\alpha_g'\ln{(2 \xi)})}
=c_g H^g(2\xi,0,t)
\label{eq:gluon-gpd}
\ee
provided $xg(x)\sim x^{1-\delta_g(t)}$. The coefficent $c_g$ is the skewing effect \ci{martin-ryskin}
which amounts to about $1.2$ in the double-distribution ansatz. From Eq.~\req{eq:gluon-gpd}
it follows that for small skewness and large $W$, the imaginary part of the scattering amplitude 
behaves as
\be
Im {\cal M}^V_{0+,0+} \sim W^{2\alpha_g(t)}
\ee
with $\alpha_g=1+\delta_g+\alpha'_gt$. From analyticity of the scattering amplitude it follows
that the real part has the same energy dependence for large $W$. Hence, the integrated
cross section behaves as
\be
\sigma_L \= \int_{t_{\rm max}}^{t_{\rm min}}dt \frac{d\sigma}{dt} \sim W^{4(\alpha_g(t_{\rm min})-1)}
\label{eq:int}
\ee
For HERA kinematics, $t_{\rm min}$ is
approximately zero. Analogous results hold for the small $\xi$ behavior of the quark GPDs.
In the handbag approach advocated for in~\cite{GK3}, the helicity amplitudes
    for transverse photons $M_{+\pm,+\pm}$ are also governed by $H^g$ at large
    $W$ and small skewness. Therefore, the above considerations and in 
    particular Eq.~\req{eq:int}
    also hold for the transverse cross section of light vector-meson
    production. Since the production of Quarkonia is also under control
    of $H^g$~\cite{Ivanov:2004vd}, its cross section also follows Eq.~\req{eq:int}.
    In Eq.~\req{eq:gluon-gpd} we see  the close connection to the leading-log approximation invented by 
Brodsky et al. \ci{brodsky} which is based on the gluon density. The difference to the handbag approach
is the skewing effect $c_g$ in Eq.~\req{eq:gluon-gpd}. The sea quarks are usually neglected in 
applications of this model.
Many variants of the leading-log approximation or color dipole
model can be found in the literature. They mainly
differ in the treatment of the subprocess $\gamma^*g\to Vg$, see e.g. 
\ci{nemchik-nikolaev-predazzi-zakharov,frankfurt-koepf-strikman,ivanov-kirchner}. 
It is also been used in \ci{VGG98}.

According to Eq.~\req{eq:int}, for vector meson production at large $W$ and small
$x_B$ the cross section behaves as $\sigma_L \sim W^{4\delta_g}$, where the power 
$\delta_g$ can be determined by experiment. The results of this 
determination are shown in the bottom panel of Fig~\ref{fig:delta} (note that $\delta_2=4\delta_g$) 
and compared to the soft Pomeron intercept~\cite{Donnachie:1999yb}. 
We see that $\delta_g$ tends to a value below that of the soft Pomeron 
intercept of 0.08. 
Despite the large errors, $\delta_g$ seems to follow a universal
    behavior $\delta_g=A_g + B_g\ln(\mu_V^2)$ which we regard as a strong indication 
    of a common underlying dynamical mechanism for light and heavy meson
    production. A realization of such a mechanism is the handbag approach
    with a dominant contribution from $H^g$.
The increase of 
$\delta_g$ with $Q^2$ reflects the increase of the gluon density for $x\to 0$ 
due to the evolution. For light vector mesons, the situation is somewhat more
complicated due to a substantial contribution from sea quarks~\ci{Meskauskas:2011aa,GK2}.
Due to quark-gluon mixing under evolution, the sea quarks, strictly 
speaking the flavor-singlet combination of quarks, is however expected to 
behave similar to the gluon and its contribution increases with approximately 
the same power of $W$ as that of the gluons.



In the intermediate energy region, for $W$ between 3 and 8 GeV, the valence quark GPDs
become important, see Fig.\ \ref{fig:vgggk}. Their contributions have to be added coherently
to the gluon and sea contributions resulting in interference effects. Since the valence-quark
PDFs also behave as a power of $x$ at low $x$ with a power $\alpha_q$ that is smaller than $\alpha_g$
at small $-t$, it is obvious that the valence-quark contribution dies off quickly with increasing
$W$ at fixed $Q^2$. For low $W$, say below about 4 GeV, the cross section tends to increase
strongly with decreasing $W$ (see Fig.~\ref{fig:vmwdep}). This is reflected
by the  negative $\delta_1$ values in Fig.~\ref{fig:delta}.
A similar strong increase is also seen for $\omega$
production but not for the $\phi$ channel. While the GK approach \ci{GK2} works well for 
$W\gsim 4\,\gev$ it fails at smaller energies (Fig.\ \ref{fig:vgggk}). 
This is also the case for other handbag analyses~\cite{VGG98,Kumericki:2011zc}.
The discrepancy between theory and data is more than one order 
of magnitude in the low $W$ region and the reason for this striking failure 
is not known today. One should note that for small $W$ at 
fixed $Q^2$ skewness is large and consequently $-t_{\rm min}$ (for instance at $W=2\,\gev$ 
and $Q^2\=4\,\gev^2$ $t_{\rm min} \simeq -0.9\,\gev^2$). Thus, in this kinematical region the
GPDs have to be extrapolated to rather large $-t$. Since the GPDs in the present parametrization
with the Regge-like profile function of Eq.~\req{eq:theory-regge-profile} are monotonously falling
functions of $-t$ it is evident that the predicted cross section becomes small at low
$W$ at fixed $Q^2$.

\begin{figure}[h!]
\begin{center}
\epsfxsize=10. cm
\epsfysize=10. cm
\centerline{\epsffile{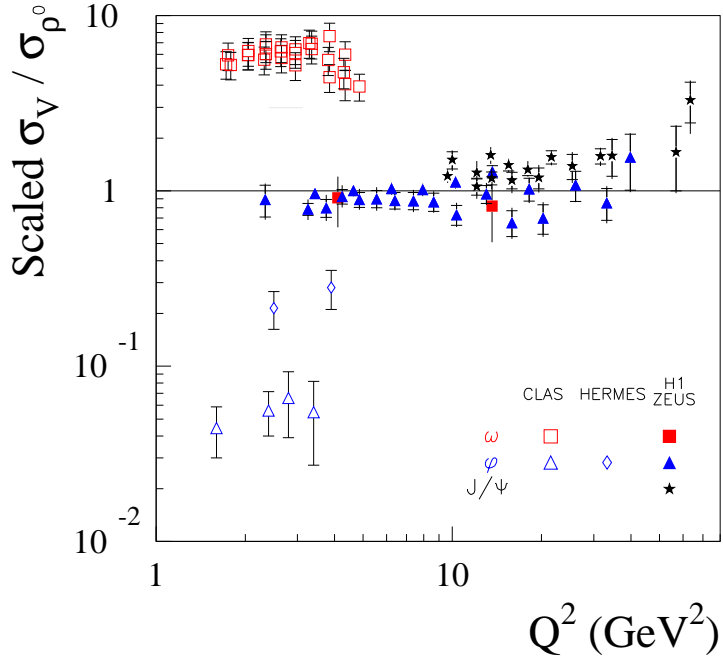}}
\caption{Ratio of the cross sections:  
$9\times\sigma_\omega/\sigma_{\rho^0}$, $9/2\times\sigma_\phi/\sigma_{\rho^0}$ 
and $9/8\times\sigma_{J/\Psi}/\sigma_{\rho^0}$ as a 
function of $Q^2$ for different $W$'s.}
\label{fig:ratio} 
\end{center}
\end{figure}

     
\subsection{Ratios of vector mesons cross sections}
     
     In search of signatures of the handbag mechanism,
     it is also instructive to look at the ratios of vector meson
      cross sections. In the large $W$/small skewness region,
      all vector meson channels are dominated by contributions from
      the gluon GPD $H^g$. Thus, up to wave function  effects, the cross sections
      differ only by the charge content of the vector mesons. Hence,
      the ratio of cross section is predicted to be:
      $\sigma_{\rho^0}/\sigma_\omega/\sigma_\phi/\sigma_{J/\Psi}=9/1/2/8$.
      In order to check these relations,
       we plot in Fig.~\ref{fig:ratio}, the ratios of the 
      cross sections: $\sigma_{V}(W,Q^2)/\sigma_{\rho^0}(W,\mu_V^2)$,
      scaled by the corresponding charge ratio. 
     Fig.~\ref{fig:ratio} shows that the experimental cross section ratios 
     are in good agreement with expectation for $W$ of about 100 GeV. 
     The data show a moderate dependence on $Q^2$. A possible explanation for this is 
     that, in addition to the gluon contribution, there is a contribution 
     from the quark sea (see Fig.~\ref{fig:vgggk}). A flavor symmetric 
     sea would also contribute according
     to the charge content. However at low scale the sea is not
     flavor symmetric. Strange and antistrange quarks are less
     abundant than the other light quarks. This effect leads to
     a mild logarithmic increase (from evolution) of the 
     $\sigma_\phi/\sigma_{\rho^0}$
     ratio. An analogous effect holds for the $J/\Psi$ since charm
     quarks are strongly suppressed at low scales but contribute
     at large ones. 

     Turning to the energy domain of $W$=4 to 5 GeV, such as available at 
     the HERMES experiment, 
     valence quarks of the proton also contribute to $\rho^0$ production, 
     about 50\% according to Fig.~\ref{fig:vgggk},
     but not to $\phi$ production. Thus, the ratio $\sigma_\phi/\sigma_{\rho^0}$
     becomes very small as is seen in Fig.~\ref{fig:ratio}. 
     For $W$ between 2 and 4 GeV, characteristic of the CLAS experiment,
     the $\rho^0$ cross section becomes very large
     while the $\phi$ cross section continues
     to decrease so that the ratio $\sigma_\phi/\sigma_{\rho^0}$ 
     gets smaller again. At these energies, the $\omega$ cross section 
     is very close to the $\rho^0$ cross section (see Fig.~\ref{fig:vmwdep})
     so that the ratio $9\times\sigma_\omega/\sigma_{\rho^0}$ plotted
     in Fig.~\ref{fig:ratio} is larger than one.
     Since the increase of the $\rho^0$
     (and $\omega$) cross section between $W$=2 and 4 GeV is not 
     understood, it is not possible at this time to explain
     and predict the values of the cross section ratios at CLAS
     kinematics.

\vskip 1.cm

\subsection{Slopes of differential cross section}

\begin{figure}[h!]
\begin{center}
\includegraphics[width=0.45\tw]{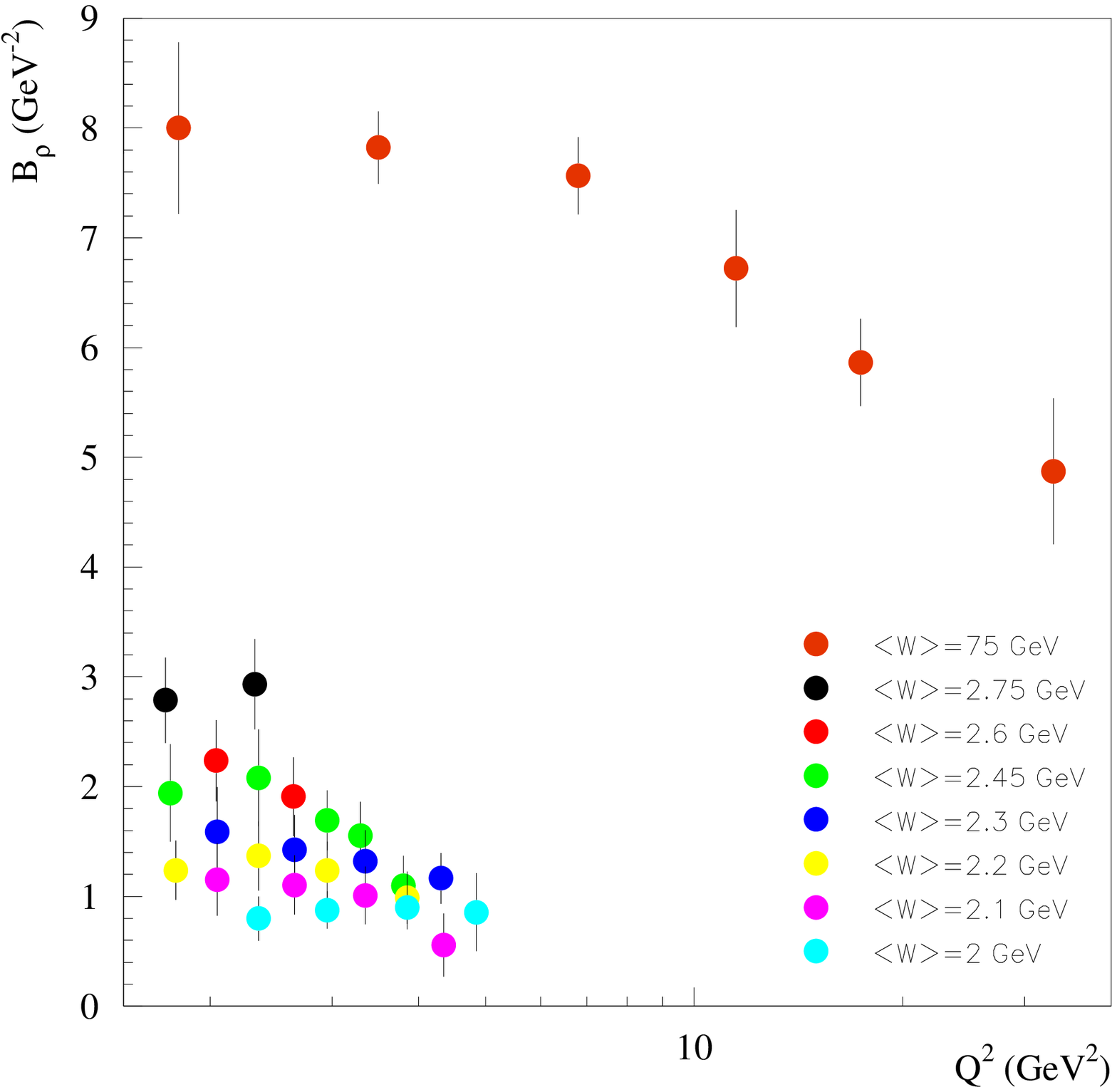}
\includegraphics[width=0.45\tw]{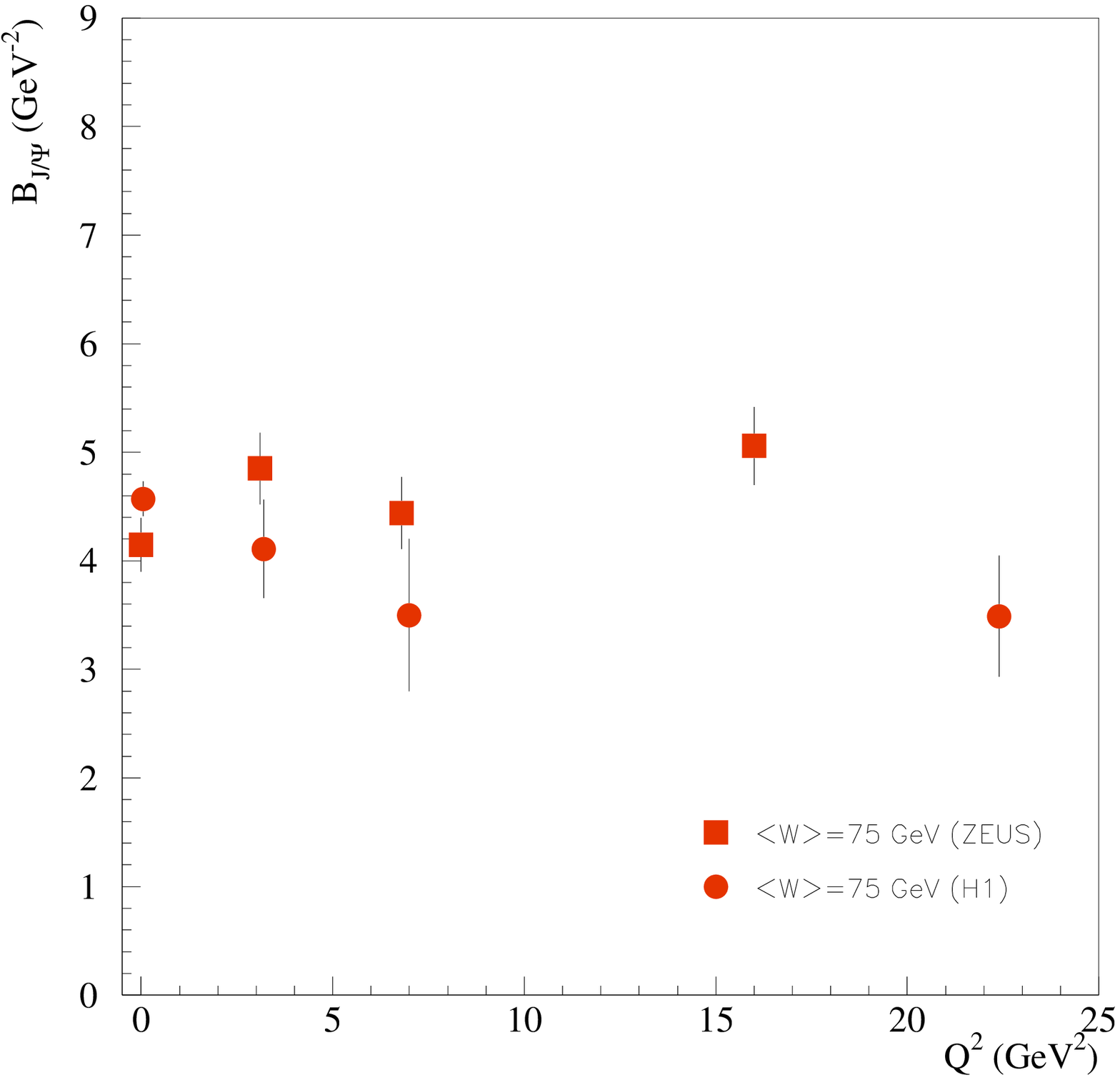}
\caption{$B$-slope parameter as a function $Q^2$. Left: for the $\rho^0$,
at different $W$'s. Right: for the $J/\Psi$ at $W$=75 GeV.}
\label{fig:bslope} 
\end{center}
\end{figure}

Regarding the $t$-dependence of the cross sections, Fig.~\ref{fig:bslope} shows 
the $Q^2$-dependence of the $B$ parameter from
the fit of the $\rho^0$ and $J/\Psi$ differential cross section 
with the function $Ae^{Bt}$.
One observes several features. 
The $\rho^0$ data at fixed $Q^2$ show an increase of $B$ with increasing $W$. 
And, at fixed $W$, concentrating on the large $W$ data (75 GeV), 
comparing the $\rho^0$ and $J/\Psi$ $Q^2$-dependence, one can deduce
a convergence of the $B$ parameter for $Q^2\approx$ 10 GeV$^2$
around 4 GeV$^{-2}$.

As we already mentioned the high-energy $J/\Psi$ photo- and leptoproduction
    amplitude is dominantly imaginary and under control of the GPD $H^g$
    at small skewness (for $J/\Psi$ production, the skewness is given by
    $\xi=(M_{J/\Psi}^2+Q^2)/(W^2+Q^2)/2$). As shown in~\cite{Strikman:2009bd},
    the slope parameter of the differential cross section is related
    to the average impact parameter of the gluon distribution at given
    $x=2 \xi$ by (see Eq.~\req{eq:impact} with the quark distribution
   replaced by the corresponding gluon one):

\be
   \langle b^2 \rangle^g_{x=2\xi}= \frac{\int d^2b b^2 g(x,b)}{\int d^2b g(x,b)}=2 B_{J/\Psi}\,.
\ee

    In deriving this result one has to assume that $H^g(\xi,\xi,t)$ is proportional
    to $H^g(2\xi,0,t)$ which for instance is the case in the double-distribution
    parametrization (Eq.~\req{eq:gluon-gpd}). According to Eq.~\req{eq:drho} 
    and with regard to the fact that
    $x<< 1$,  $\langle b^2 \rangle^g_{x=2\xi}$ can be viewed as the transverse
    size of the nucleon's gluon distribution at $x=M^2_{J/\Psi}/W^2$. From the
    data shown in Fig.~\ref{fig:bslope}, one finds 
    $d_g(2\xi)=\sqrt{ \langle b^2 \rangle^g_{x=2\xi}}\simeq 0.55 - 0.63$ fm at 
    $x \simeq 10^{-3}$ (at the scale $M^2_{J/\Psi}$) 
    which is substantially smaller than the average size 
    of the valence-quark distribution shown in Fig.~\ref{fig:theory-geometry}
    which holds at the scale 4 GeV$^2$. The gluonic size 
    may increase logarithmically with decreasing $x$~\cite{Strikman:2009bd}. 
    The
    interpretation of the $\rho^0$ slope parameter is more difficult. Quarks
    also contribute in addition to the gluon. For $Q^2$ larger than about 
    10 GeV$^2$ and high energies $B_\rho^0$ approaches $B_{J/\Psi}$ indicating the
    increasing importance of the gluonic contribution.

\vskip 2.cm

\section{Review of Pseudoscalar Mesons data and interpretation}
\label{sec:meson_pole}

\subsection{The pion pole}

\begin{figure}[h!]
\begin{center}
\includegraphics[width=0.7\tw]{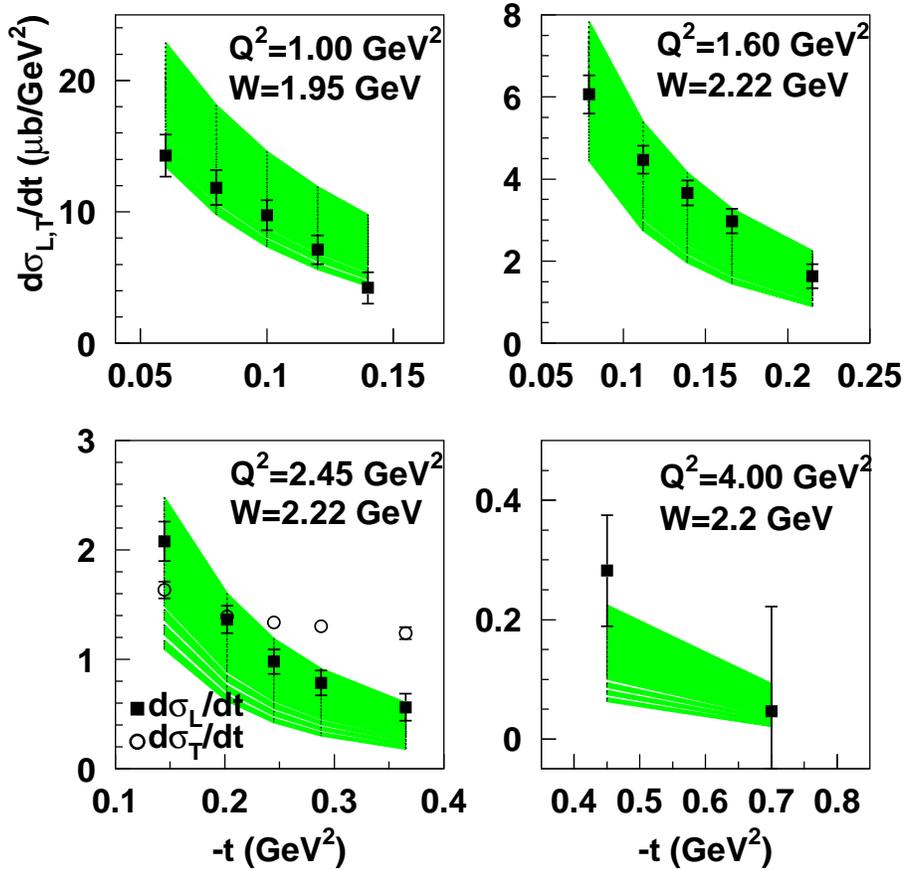}
\caption{World precision $\pi^+$ electroproduction longitudinal cross sections at two values of $W$=2.2, 1.95 GeV and values of $Q^2$ ranging from 1.0 to 4.0 GeV$^2$. The data were obtained at 6 GeV Jefferson Lab using the focusing spectrometers in Hall C~\cite{Tad06,Horn06,Horn08}. The green band indicates the calculated pole contributions according to Eq.~\req{eqn-pole} for a range of values of the parameters $\Lambda_N$=0.4-0.6 GeV and $g_{\pi NN}$=13.1-13.5. Note that the values of $W$ and $Q^2$ listed in the figures are the overall central values. Each $t$ bin has its own bin-centered $W$ and $Q^2$ values.}
\label{fig:pip_pole_sep} 
\end{center}
\end{figure}

The one-meson exchange plays an important role in leptoproduction of pseudo-scalar mesons. Its dominance in the longitudinal cross section is required for the extraction of the electromagnetic meson form factor from electroproduction data~\cite{Horn06,Hub08}. As shown in Ref.~\cite{Man99} the pion pole contribution is also part of the GPD $\widetilde E$. Working out its contribution to the longitudinal cross section to leading-twist accuracy one finds, 
\begin{equation}
\frac{d \sigma^{pole}}{dt} (\gamma^*_L \rightarrow \pi^+) = \frac{1}{\kappa} \frac{-t}{(t-m^2_\pi)^2} Q^2 \rho^2_{\pi \pi}
\label{eqn-pole}
\end{equation}
where $\rho_{\pi \pi}$=$\sqrt{2} e_0 F_\pi(Q^2) g_{\pi NN} F_{\pi NN}(t)$, $\kappa$ is a phase space factor, $F_{\pi N N}$=$\frac{\Lambda^2_N - m^2_\pi} {\Lambda^2_N -t}$ is the form factor that describes the $t$ dependence of the pion-nucleon coupling, and $F_{\pi}(Q^2)$ is the perturbative contribution to the electromagnetic form factor of the pion. The latter amounts to only 30-50\% of the experimental value of $F_{\pi}(Q^2)$=$\frac{1}{1+Q^2/\Lambda^2_{\pi}}$ ($\Lambda^2_{\pi}$=0.502 GeV$^2$) measured in the same process as we are discussing here. Using Eq.~\req{eqn-pole} with the perturbative form factor underestimates the cross section data by an order of magnitude at low $-t$. In Refs.~\cite{Kroll10,Kroll11} it has therefore been suggested to use Eq.~\req{eqn-pole} with the experimental value of $F_{\pi}(Q^2)$, which is equivalent to evaluating the pion-pole contribution as an one-pion exchange term. Working out Eq.~\req{eqn-pole} in this way one obtains the results shown in Fig.~\ref{fig:pip_pole_sep}. The bands indicate the calculated range of values for the parameters $\Lambda_N$=0.4-0.6 GeV and $g_{\pi NN}$=13.1-13.5.

Eq.~\req{eqn-pole} represents the pion pole contribution which is a special feature of the longitudinal pion electroproduction cross section. The pion pole includes a factor $-t$/$(t-m^2_\pi)^2$, which is zero at $t$=0 and reaches a maximum at $t$=$-m^2_\pi$. The first value is unphysical since forward scattering occurs at $t_{min}$=$-4m^2 \xi^2/(1-\xi^2)$ while the second can be reached in experiments for $\xi \sim m_{\pi}/2m$ . The value of the longitudinal $\pi^+$ cross sections in Fig.~\ref{fig:pip_pole_sep} is largest at small $-t<$0.3 GeV$^2$ and falls off rapidly with increasing values of $t$. The data thus suggest a dominant pion pole in the longitudinal $\pi^+$ cross section at values of $-t<$0.3 GeV$^2$. The dominance of the pion pole in the longitudinal cross section and its characteristic $t$ dependence allows for extractions of the electromagnetic pion form factor from these data. The longitudinal cross section at $W$=2.2 GeV is in good agreement with the calculation of the pion pole contribution shown in Eq.~\req{eqn-pole}. However, the pion pole calculation does not seem to describe the $Q^2$ dependence of the data very well at the lower value of $W$=1.95 GeV. The data at central values of $Q^2$=0.60, 0.75 (and 1.00) GeV$^2$ and $W$=1.95 GeV are systematically lower than the calculation. Adjusting the parametrization of the electromagnetic pion form factor to the data improves the overall agreement between data and calculation at the lower $Q^2$ settings, but does not provide an improvement of the overall description of the data set.

Unseparated $\pi^+$ data are available over a larger kinematic range in $t$ and $Q^2$. It is thus interesting to see if one can use these data to obtain additional information on the importance of the pion pole, and if this information would have an impact on pion form factor extractions. Two extensive data sets on $\pi^+$ productions are available. One was obtained at HERMES~\cite{Air08} and another one with the CLAS at 6 GeV Jefferson Lab~\cite{Park13}.

\begin{figure}[h!]
\begin{center}
\includegraphics[width=0.7\tw]{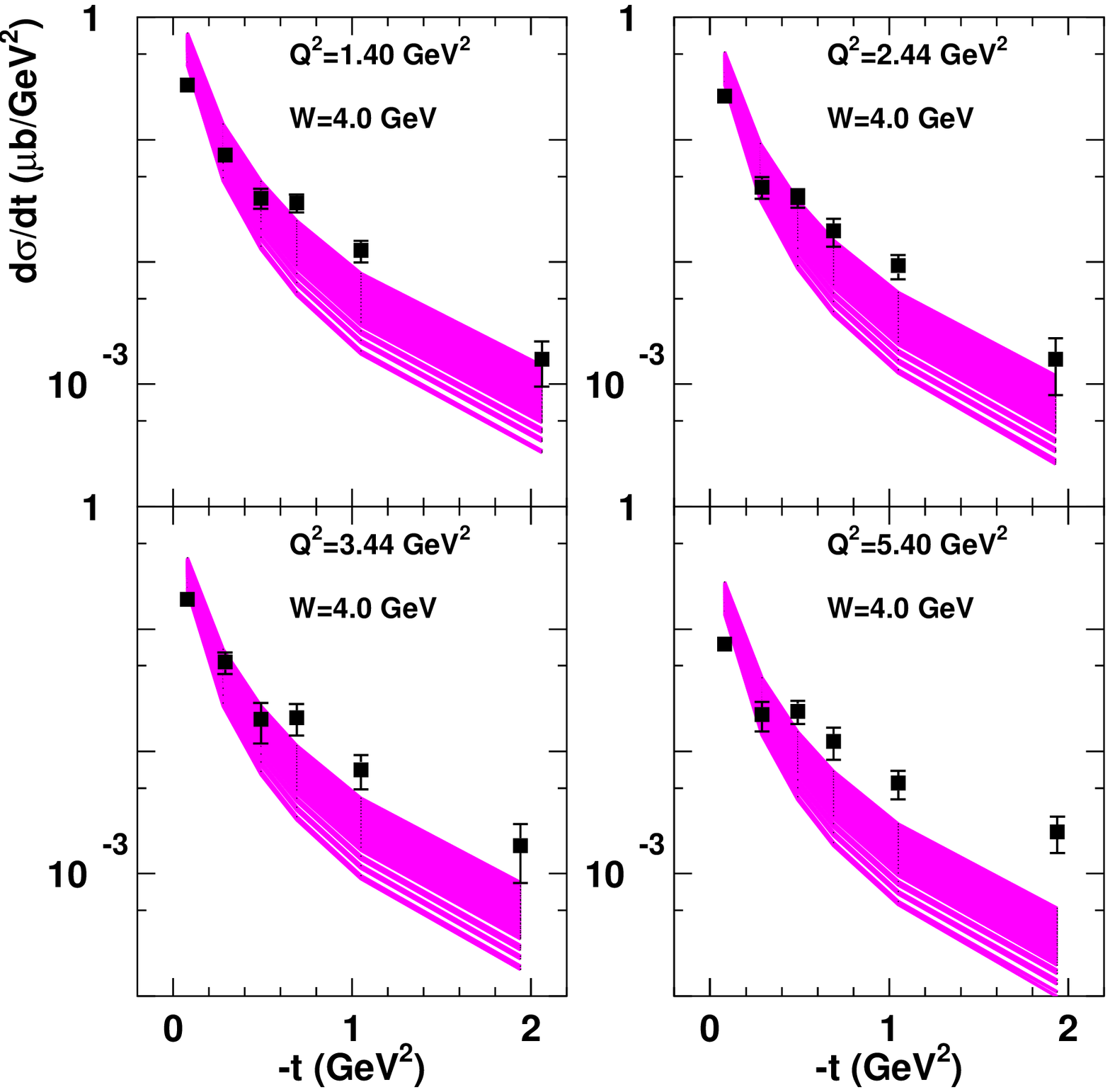}
\caption{Unseparated $\pi^+$ electroproduction cross sections from HERMES.
The bands shown are as in Fig.\ref{fig:pip_pole_sep}.}
\label{fig:pi+_pole_unsep_smallt}       
\end{center}
\end{figure}

According to Fig.~\ref{fig:pi+_pole_unsep_smallt}, Eq.~\req{eqn-pole} also describes the HERMES data at values of $-t<$0.3 GeV$^2$. This implies that the monopole parametrization of the pion form factor with $\Lambda_{\pi}$ based on pion form factor precision measurements approximately holds up to $Q^2$=5.4 GeV$^2$. At large $-t >$ 0.6 GeV$^2$, where one is farther away from the pion pole, the data deviate from the prediction by up to three sigma. The large role of the pion pole at small $-t <$ 0.3 GeV$^2$ in the unseparated $\pi^+$ cross section is interesting. However, this observation would not be sufficient to make a precise extraction of the electromagnetic pion form factor from the data. Without an explicit L/T separation it is not clear what fraction of the cross section is due to longitudinal photons and what the contribution of the pole to it is in these kinematics. 

\begin{figure}[h!]
\begin{center}
  \includegraphics[width=0.7\tw]{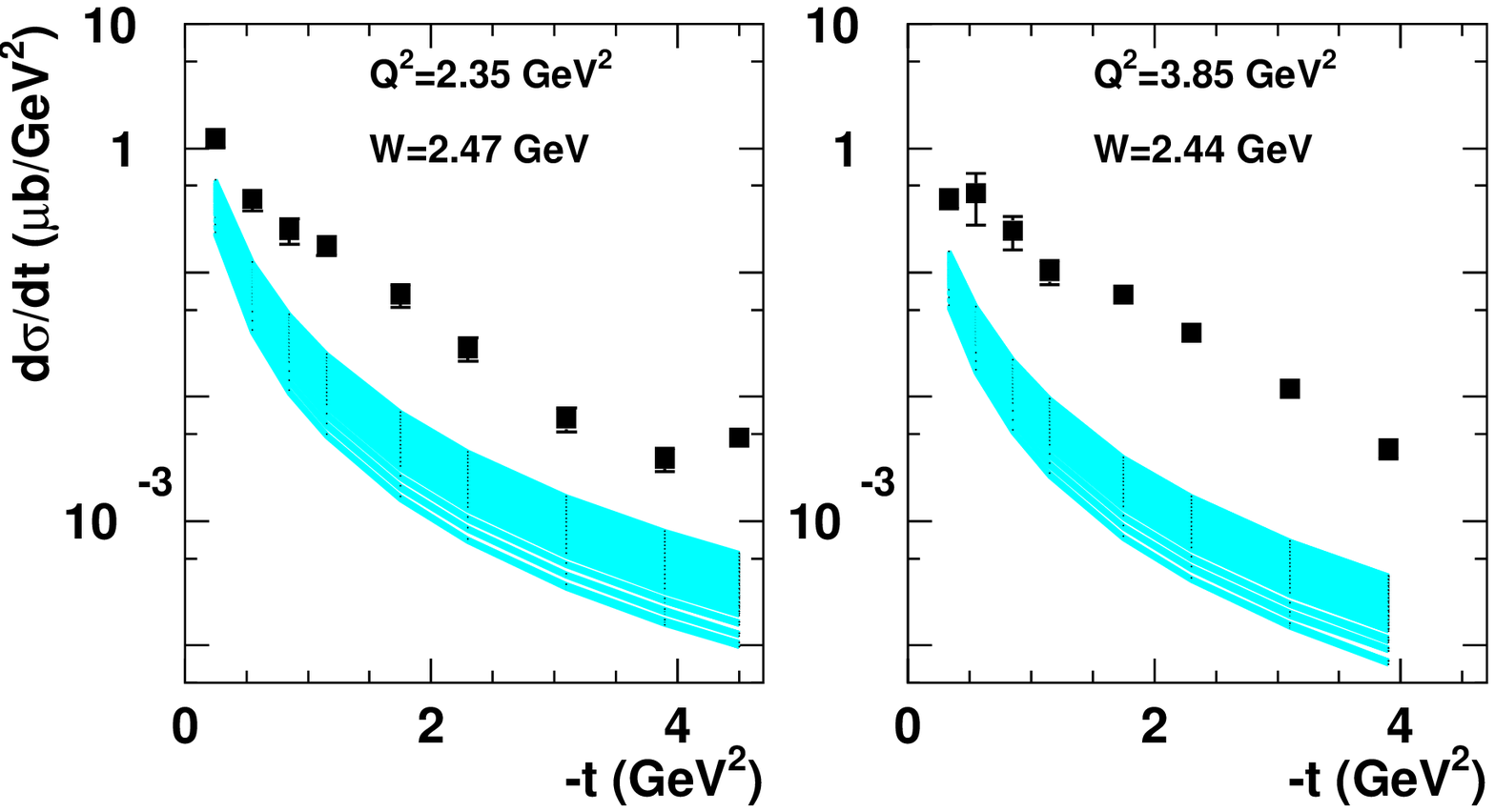}
\vskip -5.cm
\caption{Unseparated $\pi^+$ electroproduction
cross sections from 6 GeV JLab CLAS at values of $W$ of 2.47 GeV and 2.44 GeV.
The bands shown are as in Fig.\ref{fig:pip_pole_sep}.}
\label{fig:pi+_pole_unsep_larget}       
\end{center}
\end{figure}

As an illustration of effects contributing at larger values of $-t$ Fig.~\ref{fig:pi+_pole_unsep_larget} shows a comparison of the calculated pole contribution and the unseparated $\pi^+$ cross section data obtained with CLAS at Jefferson Lab 6 GeV at values of $Q^2$=2.35 GeV$^2$ and 3.85 GeV$^2$ and $W$ of about 2.5 GeV. For values of $-t$ $\sim$0.3 GeV$^2$ the calculation underpredicts the data at similar values of $Q^2$ as compared to Fig.~\ref{fig:pi+_pole_unsep_smallt} suggesting a $W$ dependence of other/non-pole contributions. Here, the tail of the pole contribution may also be competing with the background, which could be due to both longitudinal and transverse photons. To disentangle the longitudinal and transverse contributions a full separation of the cross section is needed. For a rough estimate of the transverse contribution one may compare the separated longitudinal and transverse cross sections from Fig.~\ref{fig:pip_pole_sep}. There, the contribution of the transverse cross section can be up to 90\% for $-t \sim$ 0.3 GeV$^2$.  Overall in the large $-t$ range, the calculated pion pole contribution is systematically lower than the data at both $Q^2$ settings, which may be expected since the parametrization used in Eq.~\req{eqn-pole} is only reasonable near the pion pole. 

\begin{figure}[h!]
 \begin{center}
 \includegraphics[width=0.7\tw]{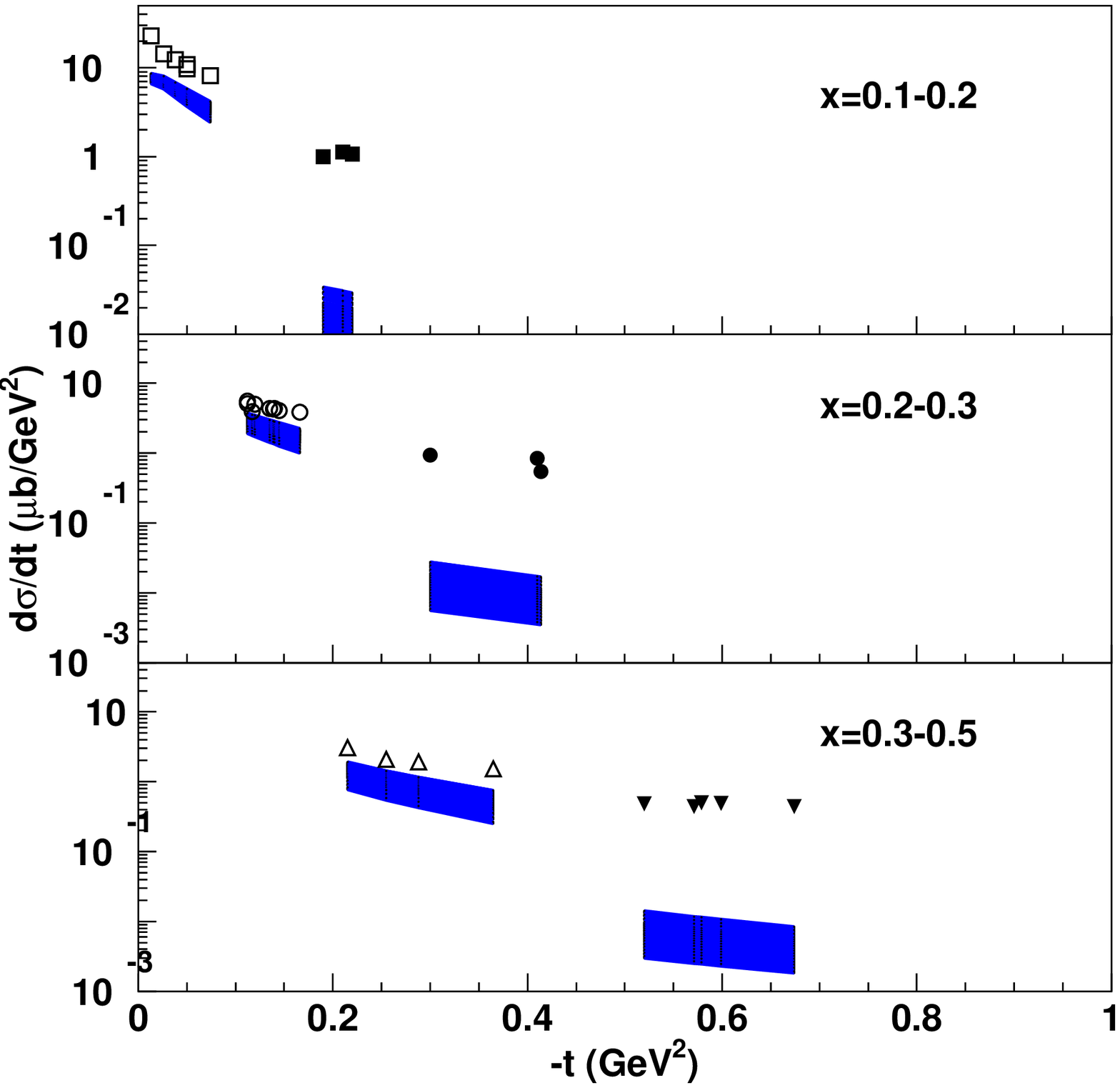}
\caption{Unseparated pion and kaon electroproduction cross sections scaled to fixed values of $Q^2$=1.6 GeV$^2$ and $W$=2.2 GeV. World pion and kaon data include~\cite{Air08,Park13,Tad06,Horn06,Horn08,Brown71,Brown71b,Beb76,Beb78,Bra76,Ack78} and ~\cite{Brown72,Beb76k,Beb77,Aze75,Bra79,Amb07,Car03,Car08,Car13,Moh03,Com10}. The open symbols denote exclusive $\pi^+$ and the filled symbols $K^+$ electroproduction data. Here, the data are binned in three $x$ bins. The blue bands denote calculations of the pole contribution according to Eq.~\req{eqn-pole}.}
\label{fig:pi+_k+_pole_unsep}       
\end{center}
\end{figure}

To study the meson pole contribution as a function of the longitudinal momentum fraction $x_B$ it is of interest to create a super set of data from different experiments at a fixed value of $W$ and $Q^2$ and to analyze its dependence on $t$. Analyzing both kaons and pions in this way allows for a comparison of the relative importance of the pole in each channel. While the contribution of the meson pole in the longitudinal $\pi^+$ cross section has been shown to be dominant allowing for pion form factor extractions~\cite{Hub08}, its role in the longitudinal $K^+$ cross section at small values of $t$ remains to be shown experimentally. Fig.~\ref{fig:pi+_k+_pole_unsep} shows unseparated $\pi^+$ and $K^+$ cross sections binned in three bins of $x_B$ and scaled to a common value of $W$=2.2 GeV and $Q^2$=1.6 GeV$^2$. The shaded areas denote the results of calculations of the meson pole contribution to the longitudinal cross section. For both the pion and the kaon pole calculation the formalism shown in Eq.~\req{eqn-pole} is used. The only changes in the calculation of the kaon pole contribution are the meson mass and the values of $g_{K \Lambda N}$ and $\Lambda_N$. The agreement of the calculation with the pion data seems to be as expected based on the results of the previous figures. There seems to be some small dependence on $x_B$. The kaon pole calculations underpredict the data by a factor of $\sim$100 for all $x_B$ bins, which seems largely due to the larger kaon mass~\cite{Horn12}. It should be noted that without the high energy and small angle capabilities offered by new facilities like the 12 GeV Jefferson Lab most of the $K^+$ data shown in Fig.~\ref{fig:pi+_k+_pole_unsep} were obtained at relatively large values of $-t>$0.3 GeV$^2$ and relatively small values of $W<$2 GeV. These data are always far away from the kaon pole, and thus its contribution to the longitudinal $K^+$ cross section is not dominant. However, a small maximum is expected in kaon production at small values of $-t$~\cite{Kroll10,Kroll11}. This needs to be verified experimentally. Measurements at the 12 GeV Jefferson Lab~\cite{E12-09-011} at lower values of $-t$ are aimed at providing such data allowing for the interpretation of the kaon pole contribution. These data will provide a full separation of the kaon cross section is required allowing for an more detailed analysis of the role of the kaon pole and the relative contribution of longitudinal and transverse photons. 

The relative contribution of longitudinal and transverse terms to the meson cross section and their $t$ and $Q^2$ dependencies are of interest in evaluating the potential of probing the nucleon's transverse spatial structure through meson production. In general, only if experimental evidence for the leading-twist behavior can be shown one can be confident about a handbag formalism. One of the most stringent experimental tests is the $Q^2$ dependence of the longitudinal meson cross section. The measurement of the fully separated longitudinal and transverse contributions to the cross section at values of $x_B$ experimentally accessible at, e.g., JLab 12 GeV, is essential for the interpretation of the data. Such data are also needed for measurements of the exclusive pion form factor and its description as discussed in the beginning of this section. In the regime where the leading-twist formalism is applicable $\sigma_L$ is predicted to scale as Q$^{-6}$, the transverse cross section is expected to scale as $\sigma_T$ $\sim$ Q$^{-8}$ and $\sigma_L >> \sigma_T$.

\begin{figure}[h!]
 \begin{center}
  \includegraphics[width=0.55\tw]{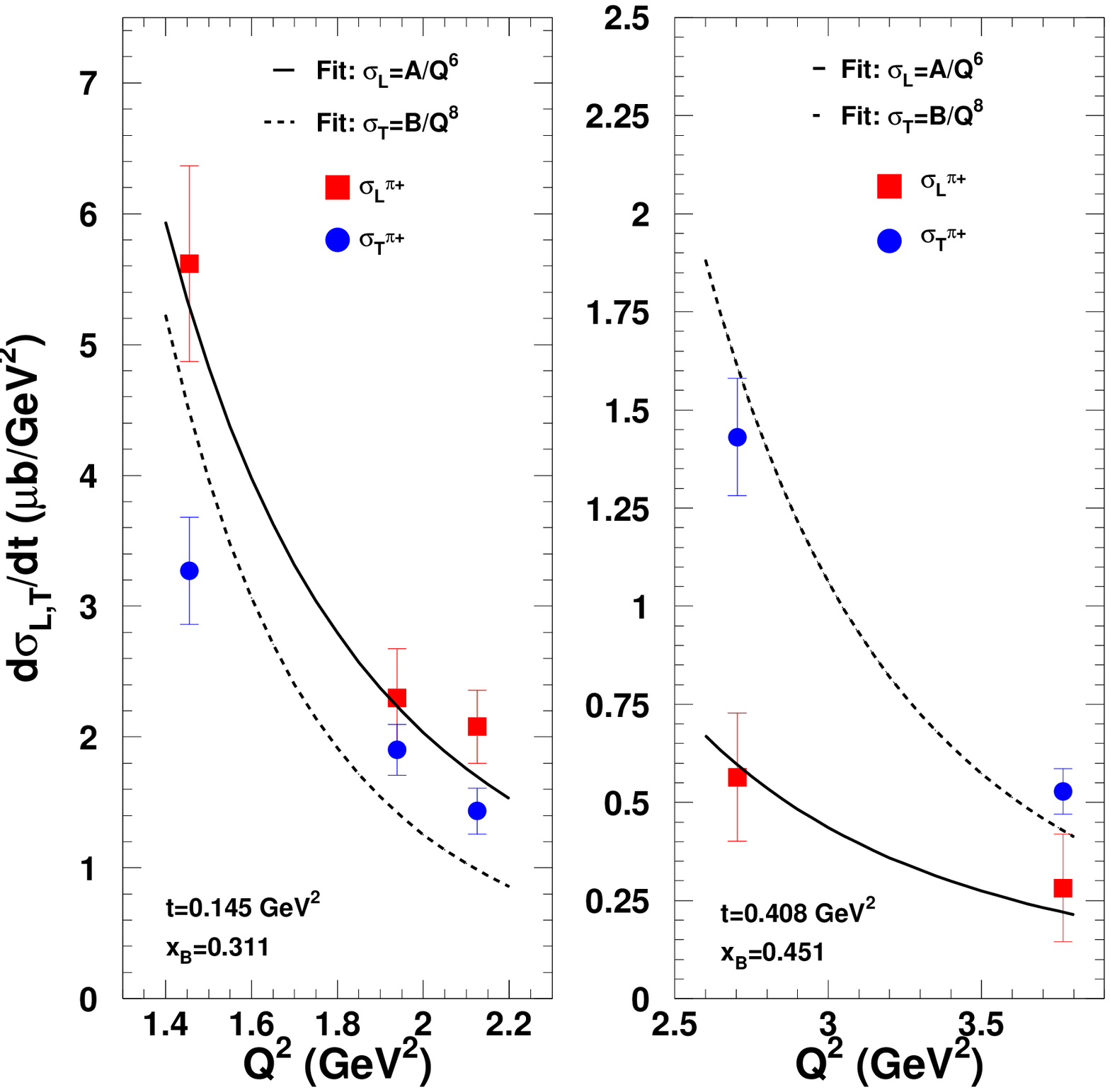}
\caption{The longitudinal and transverse $\pi^+$ electroproduction 
cross sections at fixed values of $x_B$ and $-t$. The solid line shows a fit to the $\sigma_L$ 
of the form $A/Q^6$ and the dashed line a fit to $\sigma_T$ of the form $B/Q^8$. The data are from Refs.~\cite{Horn06,Tad06,Horn08}}
\label{fig:sep_pi+_fact_alt}       
\end{center}
\end{figure}

The leading-twist, lowest order calculation of the $\pi^+$ longitudinal cross section underpredicts the data by an order of magnitude. This implies that the data are not in the region where the leading-twist result applies. That current experimental data are not in the region where the leading-twist result applies can be seen in Fig.~\ref{fig:sep_pi+_fact_alt} showing the $Q^2$ and $t$ dependence of the separated longitudinal and transverse $\pi^+$ cross sections. The QCD scaling prediction is fitted to, and indicated by, the solid black lines and is reasonably consistent with these data. It is clear $\sigma_T$ does not follow the scaling expectation illustrated by the dashed black lines and the magnitude is large. Regarding the $-t$ dependence, Fig.~\ref{fig:sep_pi+_fact_alt} shows that $\sigma_L > \sigma_T$  for values of $-t<$ 0.3  consistent with a dominant meson pole in this region and that $d\sigma_T > d\sigma_L$ for values of $-t>$ 0.3 GeV$^2$ providing further evidence that the leading-twist does not apply in the currently available experimental kinematics. 

The deviation of the $\pi^+$ cross sections from the leading-twist formalism may not be surprising. As discussed in section 3 for the vector meson $\rho^0$ production, the cross section can easily deviate from the anticipated $Q^2$ behavior in the scaling regime by power and logarithmic corrections. However, the measurements of the separated pseudoscalar meson cross sections and their dependencies are fundamental and important in their own right. Fully separated cross sections are essential for understanding dynamical effects in both $Q^2$ and $-t$ and interpretation of non-perturbative contributions in experimentally accessible kinematics. Such measurements of L-T separated cross sections will be enabled by the 12 GeV Jefferson Lab extending the current kinematic reach of $\pi^+$ data and including additional systems. These data will play an important role in our understanding of meson pole dominance and form factor extractions, and may provide experimental evidence allowing for interpretation of the data in the handbag formalism. In passing, we note 
that the pion pole also plays a prominent role in leptoproduction of omega mesons. In contrast to
pion production the pole term contributes to the transverse amplitudes in this case~\cite{GK8}.

\subsection{Transversity}


Recent pion cross section data~\cite{Horn06,Horn08,Air10,Bedl12,Fuch11} suggest that transversely polarized photons play an important role in charged and neutral pion electroproduction. As shown 
in Fig.~\ref{fig:pip_pole_sep} L/T separated $\pi^+$ data show a large $\sigma_T$ even at values of $Q^2$=2.5 GeV$^2$ and $-t<$0.3 GeV$^2$. At HERMES a large $sin (\phi_S)$  modulation was observed in the Fourier amplitude or transverse target spin asymmetry, $A_{UT}(sin (\phi_s))$, which does not seem to vanish in the forward direction~\cite{Air10}. The observed behavior of the $A_{UT}$ data demand a strong contribution from transverse photons. For $(t_{\rm min}-t)
 \rightarrow$ 0 the only non-vanishing contribution to this observable is $Im (M_{0-,++}^* M_{0+0+})$. Thus, the transverse amplitude $M_{0-,++}$ must be of similar strength as the asymptotically leading longitudinal amplitude $M_{0+0+}$ for $\pi^+$.

\begin{figure}[h!]
 \begin{center}
  \includegraphics[width=0.7\tw]{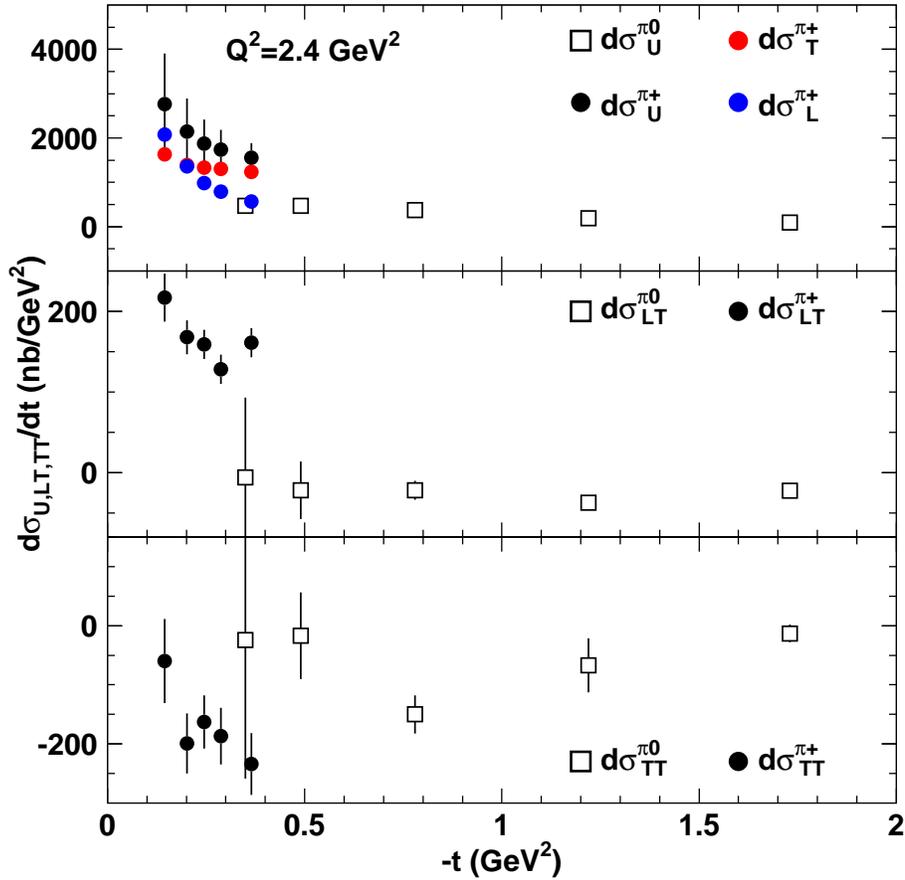}
\caption{The L-T and T-T interference and unseparated pion cross sections (separated for $\pi^+$) at a value of $Q^2$=2.4 GeV$^2$ and $x_B$ $\sim$ 0.4. The data are from Refs.~\cite{Horn06,Hub08,Bedl12}.}
\label{fig:siglt_pion_q245}       
\end{center}
\end{figure}

\begin{figure}
 \begin{center}
  \includegraphics[width=0.6\tw]{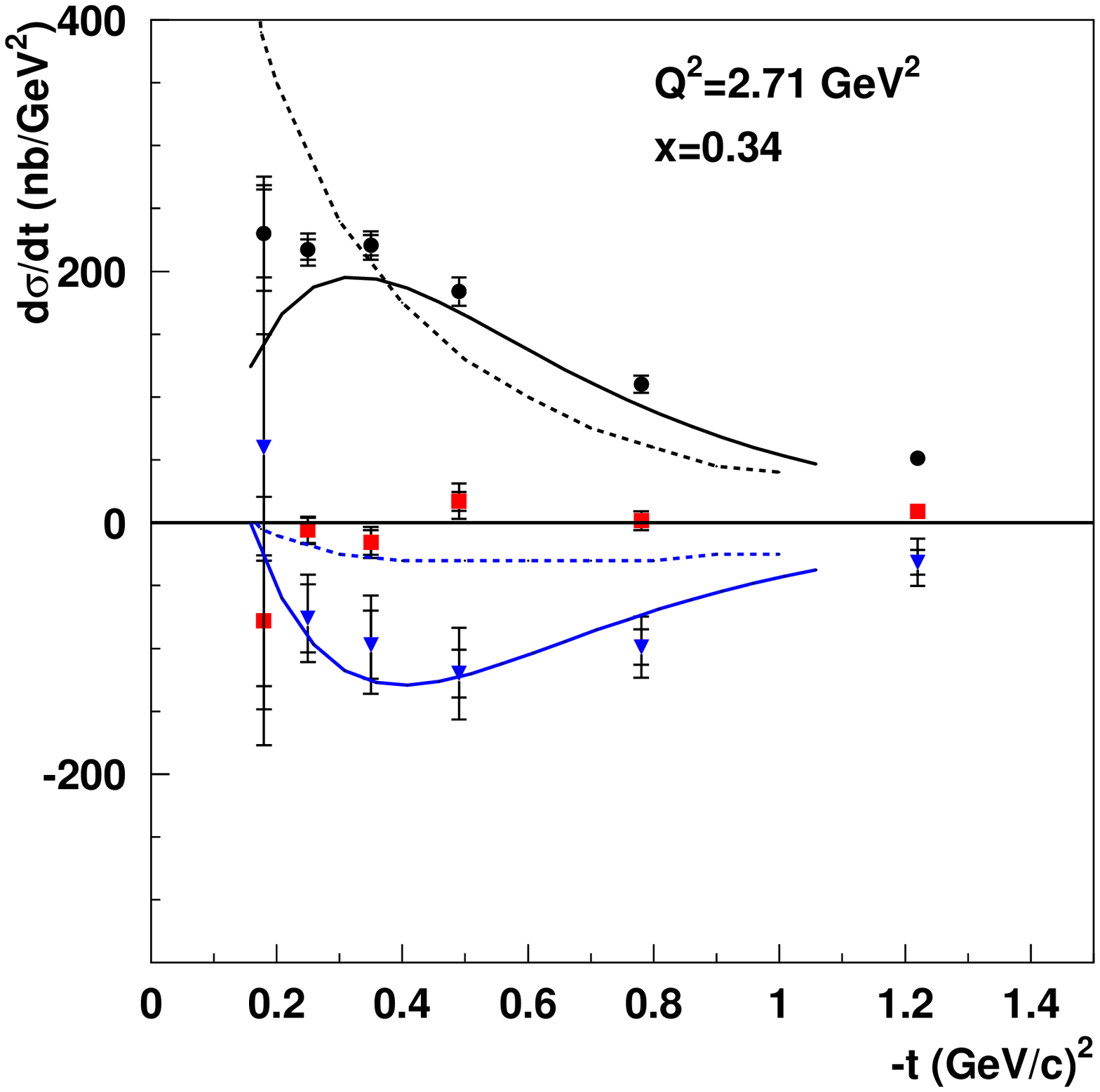}
\caption{The L-T and T-T interference and unseparated pion cross sections at a value of $Q^2$=2.7 GeV$^2$ and $x_B$ $\sim$ 0.4 from Ref.~\cite{Bedl12} compared to the two theoretical predictions of Refs.~\cite{Kroll11} (solid) and ~\cite{Ahm09,Ahm09b,Ahm09c} (dashed). Here, $t_{min}$=-0.15 GeV$^2$. The black symbols denote the unseparated cross section $\sigma_U$=$\sigma_T + \epsilon \sigma_L$, the blue symbols the T-T interference ($\sigma_{TT}$), and the red symbols the L-T interference ($\sigma_{LT}$). The inner error bars on the data points 
denote the statistical error and the outer error bars the statistical and systematic
errors combined in quadrature.}
\label{fig:siglt_pion_models}
\end{center}
\end{figure}

Fig.~\ref{fig:siglt_pion_q245} shows separated charged~\cite{Horn06,Tad06,Horn08} and unseparated neutral pion cross section data from 6 GeV Jefferson Lab~\cite{Bedl12}. For the charged pion data represented by the filled symbols in the upper panel the dominant role of the pole at values of $-t<$0.3 GeV$^2$ is evident. Further one can see the characteristic fall off of $\sigma_L$ as $t$ increases resulting in $\sigma_T$ to become larger than $\sigma_L$ at values of $-t>$0.3 GeV$^2$. The LT interference term in $\pi^+$ production is non-zero and positive while the TT interference term is negative and consistent with zero within the uncertainty over the region in $t$ shown. This behavior is consistent with that expected from angular momentum conservation shown in Fig.~\ref{fig:siglt_pion_models}. The interference cross sections $d\sigma_{LT}$ and $d\sigma_{TT}$ vanish as $\sqrt{t_{\rm min}-t}$ and 
as $(t_{\rm min}-t)$ respectively. Thus, the  interference cross sections become small for 
$t_{\rm min}-t \rightarrow$ 0. 

When looking at current neutral pion data as those presented in Fig.~\ref{fig:siglt_pion_q245}, it should be noted that these data are only available for values of $t \sim$0.2 GeV$^2$ and are not fully separated. The L-T interference term in $\pi^0$ production is small. In Fig.~\ref{fig:siglt_pion_q245} it is negative on average, but it can also be positive depending on kinematics. In general, a non-zero L-T interference term would suggest that there is a longitudinal contribution to the cross section. The TT interference term in $\pi^0$ production is large in absolute value suggesting that transverse photons play an important role in this kinematic regime.

To interpret the data including a large contribution from transverse photons in Refs.~\cite{Kroll10,Kroll11} the handbag approach is generalized to $\gamma^*_T\to M_L$ transition amplitudes. They are represented by convolutions of transversity GPDs and subprocess amplitudes calculated with a twist-3 pion wave function. The latter come along with a mass parameter $\mu_\pi$ which is given by the pion mass, $m_\pi$, enhanced by the chiral condensate 
\begin{equation}
     \mu_\pi = m_\pi^2/(m_u+m_d)
\end{equation}
by means of the divergency of the axial vector current ($m_u$ and $m_d$ denote current quark masses). The transverse amplitudes are parametrically suppressed by $\mu_\pi/Q$ as compared to the asymptotically leading longitudinal amplitudes. As we mentioned in the first section,
collinear factorization does not hold for the transverse amplitudes. However, if one allows for quark transverse momenta in the subprocess the infrared singularities disappear and $k_\perp$-factorization for the transverse amplitudes may hold.

As shown in Refs.~\cite{Kroll10,Kroll11} the dominant transversity contributions at least at small $-t$ and small skewness are
\begin{eqnarray}      
     {\cal M}_{0-,++} &=& e_0\sqrt{1-\xi^2} \int dx {\cal H}_{0-,++} H_T\\
     {\cal M}_{0+,\pm +} &=&-e_0\frac{\sqrt{t_{min}-t}}{4m}
         \int dx {\cal H}_{0-,++} \bar{E}_T\,.
\end{eqnarray}  
Parametrizing the transversity GPDs according to this formalism the trends and magnitudes of the $\pi^+$ and the interference terms of the $\pi^0$ data from JLab and HERMES are well described. Transversity GPDs in pion electroproduction have also been discussed in~\cite{Ahm09,Ahm09b,Ahm09c}.

To confirm the estimates of the contribution of transverse photons and the potential to access GPDs in meson production requires a full separation of the cross section. The trends discussed above depend on both $Q^2$ and $t$, and thus it is important to cover as large of a kinematic range as possible including the regime $-t<$0.3 GeV$^2$. 
The first L/T separated $\pi^0$ cross sections were measured Hall A at Jefferson Lab 6 GeV and are under analysis. 
These data cover a range in $Q^2$ between 1.5 and 2 GeV$^2$ and $x_B$ of 0.36. 
A larger kinematic coverage for both charged and neutral pion (and kaon) production can be achieved with approved experiments at 12 GeV Jefferson Lab~\cite{E12-13-010,E12-07-105,E12-09-011}. If experimental evidence for the dominance of $\sigma_T$ can be demonstrated to hold, one may use these data to probe transversity GPDs.

\section{Summary}


Detailed inspection of the data on hard exclusive meson
   lepto- and photoproduction reveals clear evidence for a common
   dynamical mechanism underlying these processes for which the
   handbag approach is a serious candidate. 
However, most of the currently available experimental data are not in a region in which the leading-twist result applies. Only at very small values of $x_B$ and $Q^2$ 
larger than about 50 GeV$^2$ do the data follow the leading-twist result. Nevertheless,
 detailed measurements of DVMP cross sections are important in order to understand dynamical effects and the interpretation of non-perturbative contributions in experimentally accessible kinematics. An example are the substantial contributions of transverse photons to the light meson channels, e.g., in neutral pion production. The dominance of transverse photons opens, if experimentally verified, new and unique opportunities for accessing the transversity GPDs over a large kinematic range. 

On the theory front much progress has been made by including the contributions of transverse photons in light meson production into the handbag approach. This generalized handbag approach accounts for all observed features in the data with the exception of 
$\rho^0$ and $\omega$ production below values of $W<$ 4 GeV.

Future measurements of DVMP cross sections will allow for confirming the estimates of transverse photon contributions and the potential to access GPDs in meson production, as well as to understand the remaining puzzles, e.g., the low W $\rho^0$ and $\omega$ cross section data. Such data are soon expected in the valence quark region with the 12 GeV upgraded JLab. With a beam energy of 11 GeV, a luminosity of $10^{35}$cm$^{-2}$s$^{-1}$ and 1000 hours of beam time with the CLAS12 detector for instance, 
about 100 million $\rho^0$ events are estimated to be collected (taking into account the CLAS12 acceptance). In particular, $x_B$ values from $<$0.1 up to 0.8 and
$Q^2$ values up to 12 GeV$^2$ can be reached, with L/T separation over most of the phase space
for vector mesons. 

These measurements of the vector meson cross sections in a large-acceptance setup will go hand-in-hand with precision pseudoscalar meson cross section measurements in Hall C. The heavily-shielded detector setup in a highly-focusing magnetic spectrometer with large momentum reach, rigid connection to a sturdy pivot, well-reproducible magnetic properties, and access to the highest-luminosity data 
($10^{38}$cm$^{-2}$s$^{-1}$), provide the essential factors for meaningful longitudinal-transverse cross section separations. The anticipated excellent resolution and systematic understanding (less than 2\% point-to-point) of the HMS-SHMS spectrometer pair best address the experimental requirements for this program.

Beyond the opportunities for DVMP studies in the kinematic region of the valence quarks afforded by the 12 GeV upgrade of CEBAF at JLab, the planned Electron-Ion Collider (EIC) will allow for exploring the role of the gluons and sea quarks in determining the hadron structure and properties. For example, measurements of Compton scattering as well as exclusive $\rho^0$ and $J/\Psi$  production at high $Q^2$ could 
allow one to disentangle the singlet quark and gluon GPDs, and test the QCD evolution. The EIC also provides the facilities for measurements of two mesons with a large rapidity gap between them, 
which could be another interesting avenue to probe the GPDs for transversally polarized quarks. 
Further details on the EIC and its science can be found in Chapter ``TMD and GPDs at EIC" 
in this review or Refs.~\cite{accardi,doe}.

\section*{Acknowledgments}

We are very thankful to A. Borissov, S. Goloskokov, X. Janssen, R. McNulty
for the help they kindly provided. L. F. is supported by the ``Fonds de la
Recherche Scientifique -FNRS" of Belgium. M.G. benefitted from the 
ANR-12-MONU-0008-01 ``PARTONS" contract support. T. H. is supported in part 
by NSF grant PHY-1306227.

\end{document}